\documentclass{aa}

\usepackage{graphicx}
\usepackage{txfonts}
\usepackage{color}
\usepackage{ulem}
\begin{document}

   \title{Planetesimal formation at the gas pressure bump following a migrating planet}

   \subtitle{II. Effects of dust growth}

   \author{Y. Shibaike\inst{1}
          \and
          Y. Alibert\inst{1}
          }

   \institute{Physikalisches Institut \& NCCR PlanetS, Universitaet Bern, CH-3012 Bern, Switzerland\\
              \email{yuhito.shibaike@unibe.ch}
             }

   \date{Received MM DD, 2020; accepted MM DD, 2020}

 
  \abstract
   {Planetesimal formation is still mysterious. One of the ways to form planetesimals is to invoke a gas pressure bump in a protoplanetary disc. In our previous paper, we propose a new scenario in which the piled-up dust at a gas pressure bump created by a migrating planet form planetesimals by streaming instability in a wide region of the disc as the planet migrates inward.}
   {In this work, we consider the global time evolution of dust and investigate the detailed conditions and results of the planetesimal formation in our scenario.}
   {We use a 1D grid single-sized dust evolution model, which can follow the growth of the particles by their mutual collision and their radial drift and diffusion. We calculate the time-evolution of the radial distribution of the peak mass and surface density of the dust in a gas disc perturbed by an embedded migrating planet and investigate if the dust satisfies the condition for planetesimal formation.}
   {We find that planetesimals form in a belt-like region between the snowline and the position where the planet reaches its pebble-isolation mass when the strength of turbulence is $10^{-4}\leq\alpha\leq10^{-3}$, which is broadly consistent with observed value of $\alpha$. The mechanism of the formation, streaming instability or mutual collision, depends on the timescale of the streaming instability. The total mass of planetesimals formed in this scenario also depends on $\alpha$ and is about $30-100~M_{\rm E}$ if the planetary core has already existed at the beginning and grows by gas accretion, but it decreases as the timing of the formation of the planetary core is later. We also provide simple approximate expressions of the surface density and total mass of the planetesimals and find that the total planetesimal mass strongly depends on the dust mass.}
   {We show that planetesimals form in a belt-like region by the combination of the dust pile-up at the gas pressure bump formed by a planet and its inward migration.}

   \keywords{planets and satellites: formation -- protoplanetary disks -- planet-disk interactions -- methods: numerical
               }

   \maketitle

\section{Introduction} \label{introduction}
The formation of planetesimals, kilometer-sized building blocks of planets, has been investigated for a long time, but a lot of problems still remain. Especially, the so-called ``drift barrier'' is not solved yet. Planetesimals had been considered to form by mutual collisions of dust particles in protoplanetary discs. The particles, however, suffer head wind from the gas disc rotating with a sub-Kepler speed due to the gas pressure gradient and lose their angular momentum, which forces the particles to drift toward the central star before they grow to planetesimals \citep[e.g.][]{whi72}.

One of the solutions to avoid the loss of the particles by inward drift is invoking the gas pressure bump at some location in the disc. The gas pressure gradient is null at the bump, and drifting particles pile up there \citep[e.g.][]{zhu12}. Many observations of the millimeter continuum emission from protoplanetary discs show ring and gap structures \citep[e.g.][]{ALMA2015,and18,seg20}, which are considered as the evidence of the dust pile-up at gas pressure bumps \citep{dul18}. One of the most popular mechanisms to form the ring and gap structures is the gravitational interaction with embedded planets \citep[e.g.][]{paa04}, and some observations of protoplanets in the gaps support this mechanism \citep[e.g.][]{kep18,pin19,cur22}. Such dust concentrated locations are suitable for planetesimal formation by gravitational instabilities or by mutual sticking (collision) of the dust \citep[e.g.][]{sek98}. Especially, streaming instability occurs by the accumulation of dust and makes clumps of dust, which triggers the further gravitational instability and forms planetesimals \citep[e.g.][]{you05}. \citet{sta19} shows that if dust particles at the gas pressure bump form planetesimals by streaming instability, the dust rings in multiple protoplanetary discs observed by DSHARP survey observation \citep{and18} are better explained. Many previous works also argue that planetesimals can form at the gas pressure bump created by an embedded planet \citep[e.g.][]{lyr09,ayl12a,cha13,dra19,eri20}. The planetesimals formed at the bump grow larger and parts of them are captured or scattered by the planet \citep{kob12,eri20,eri21}.

In our previous paper, \citet{shi20} (hereafter Paper 1), we proposed a new scenario in which planetesimals form by streaming instability at the gas pressure bump created by a migrating planet, resulting in planetesimal formation in a wide region of the protoplanetary disc. We developed a simple 1D Lagrangian particle model which can follow the radial distribution of fixed-sized dust in a gas disc perturbed by a migrating planet. We showed that planetesimals form in a wide region of the disc, and their total mass and formation region depend on the dust mass flux and the strength of turbulence in the disc. We also found that the surface density of formed planetesimals can be approximated by a simple equation. \citet{mil21} reproduced the observed exoKuiper belts (i.e., planetesimal belts in extrasolar systems) by this scenario with a simple grid model of the global dust evolution. The surface density profiles of the formed planetesimals in \citet{mil21} are consistent with the approximate expression by Paper 1.

In this paper, we investigate the detailed conditions and results for our planetesimal formation scenario by considering global dust evolution. We do not use the Lagrangian model developed in Paper 1 but use a grid model which can follow the time evolution of the radial profiles of the peak mass and surface density of dust particles. We assume the existence of a migrating planet (or a planetary core) carving the gas disc and investigate when and where the planetesimals form by streaming instability or by mutual collision by changing the strength of turbulence and the (poorly known) condition for streaming instability. Although this work is similar to \citet{mil21}, we do not focus on the reproduction of observations but on the detailed investigation of the phenomena of planetesimal formation. Also, we consider an earlier stage of planet formation, when the planet does not migrate in Type II migration but does in Type I migration.

In Section \ref{methods}, we explain the methods used in this work. We then show the results of the calculation depending on the timescale of streaming instability in Section \ref{results}. We also explain a case where the properties of streaming instability depends on the Stokes number of dust. In Section \ref{discussion}, we investigate the effects of change of disc properties. Furthermore, we investigate the effects of the planetary growth by gas accretion and the later formation of the planetary cores considering the shift of the migration type from the Type I to Type II and the time evolution of the disc. We also discuss the effects of the back reaction from dust to gas and the dust leak from gas pressure bumps. Finally, we conclude this work in Section \ref{conclusions}.

\section{Methods} \label{methods}
\subsection{Gas disc model} \label{gas}
First, we set a gas disc model. The unperturbed (i.e., not perturbed by a planet) gas surface density is assumed to follow a power low:
\begin{equation}
\Sigma_{\rm g,unp}=\Sigma_{\rm g,1au}\left(\dfrac{r}{\rm au}\right)^{-p},
\label{gasg}
\end{equation}
where $r$ is the distance to the star, and $\Sigma_{\rm g,1au}$ and $p$ are constants. The disc temperature (in the midplane) is
\begin{equation}
T=T_{\rm 1au}\left(\dfrac{r}{\rm au}\right)^{-q},
\label{temperature}
\end{equation}
where $T_{\rm 1au}$ and $q$ are constants. We assume the constants as $\Sigma_{\rm g,1au}=500~{\rm g~cm^{-2}}$, $T_{\rm 1au}=280~{\rm K}$, $p=1$, and $q=1/2$. This set of assumption is consistent with the ``Model A'' of Paper 1. The slope of the gas surface density is consistent with the observations of protoplanetary discs under the assumption that the dust-to-gas surface density ratio is uniform through the entire discs \citep{and10}. We set the snowline at the orbit where $T=160~{\rm K}$, which is $r_{\rm SL}=3.06~{\rm au}$ in this disc model.

We note that when the disc temperature is dominated by the viscous heating, the temperature increases as the turbulence is stronger. In this paper, however, we fix the gas surface density and temperature with changing the strength of turbulence. We investigate the cases with a hotter disc in Section \ref{dependence} and with a time-evolving gas disc in Section \ref{realistic}.

\subsection{Gap formation by a migrating planet} \label{gap}
Planets embedded in gas discs influence the discs and changes the gas structure. We assume the presence of a single planet with fixed mass $M_{\rm pl}=20~M_{\rm E}$, migrating inward from $r=50~{\rm au}$. In Section \ref{realistic}, we consider the growth of the planet by gas accretion and its later formation. The subscript ``pl'' indicates the properties of the planet and/or its location. The surface density of the local perturbed gas disc has been modeled by many previous works. We use a model provided by \citet{duf15a} in order to compare our results with the pebble-isolation mass provided by \citet{ata18} (see next paragraph), which also uses the model of \citet{duf15a}\footnote{In Section \ref{realistic}, we use the model described in Paper 1, because the model by \citet{duf15a} is not accurate when the planet is heavy.}. The perturbed gas surface density profiles is
\begin{equation}
\Sigma_{\rm g}=\Sigma_{\rm g,unp}\left\{1 - \dfrac{f(r)K/(3\pi)}{1 + f_{0}K/(3\pi)}\sqrt{r_{\rm pl}/r}\right\}
\label{Sigmag}
\end{equation}
where $r_{\rm pl}$ is the orbital radius of the embedded planet, and the parameter $f_{0}$ is fixed as $0.45$. The factor $K$ is defined as
\begin{equation}
K\equiv\left(\dfrac{M_{\rm pl}}{M_{*}}\right)^{2}\left(\dfrac{H_{\rm g,pl}}{r_{\rm pl}}\right)^{-5}\alpha^{-1},
\label{K}
\end{equation}
where $M_{*}=1M_{\odot}$ is the mass of the star, and $\alpha$ is the strength of turbulence of the gas \citep{kan15a}. We treat $\alpha$ as a constant (in space and time) and change the value as a parameter. The gas scale height (at $r_{\rm pl}$) is $H_{\rm g,pl}=c_{\rm s,pl}/\Omega_{\rm K,pl}$, where the sound speed and the Kepler frequency are $c_{\rm s,pl}=\sqrt{k_{\rm B}T_{\rm pl}/m_{\rm g}}$ and $\Omega_{\rm K,pl}=\sqrt{GM_{*}/r_{\rm pl}^{3}}$, respectively\footnote{These expressions are also valid without the subscripts.}. The Boltzmann constant and the gravitational constant are $k_{\rm B}$ and $G$, respectively. The mean molecular mass is $m_{\rm g}=3.9\times10^{-24}~{\rm g}$. The function $f(r)$, the scaled-out angular momentum flux by the shocking of the planetary wake, is
\begin{equation}
f(r)=
\begin{cases}
f_{0}, & \tau(r)<\tau_{\rm sh}, \\
f_{0}\sqrt{\tau_{\rm sh}/\tau(r)}, & \tau(r)\geq\tau_{\rm sh}.
\end{cases}
\label{dustsupply}
\end{equation}
where the shock position, $\tau_{\rm sh}$, is given as \citep{goo01}
\begin{equation}
\tau_{\rm sh}=1.89+0.53\left(\dfrac{M_{\rm pl}}{M_{*}}\right)^{-1}\left(\dfrac{H_{\rm g,pl}}{r_{\rm pl}}\right)^{3}.
\label{taush}
\end{equation}
The parameter $\tau(r)$, representing an appropriately scaled distance from the planet, is
\begin{equation}
\tau(r)=\dfrac{3}{2^{5/4}}\left(\dfrac{H_{\rm g,pl}}{r_{\rm pl}}\right)^{-5/2}\left|\int^{r/r_{\rm pl}}_{1}|s^{3/2}-1|^{3/2}s^{p/2+5q/4-11/4}ds\right|.
\label{taur}
\end{equation}

Once a gap forms around a planet, the dust particles start to pile up at the gas pressure bump, and their accretion onto the planet stops. The planet mass where the dust (pebble) accretion stops is called ``pebble-isolation mass'' \citep[e.g.][]{lam14}. \citet{ata18} found that it depends on the planet mass and the strength of turbulence,
\begin{equation}
M_{\rm PIM}=h_{\rm pl}^{3}\sqrt{37.3\alpha+0.01}\left\{1+0.2\left(\dfrac{\sqrt{\alpha}}{h_{\rm pl}}\sqrt{\dfrac{1}{{\rm St_{pl}}^{2}}+4}\right)^{0.7}\right\}M_{\rm *},
\label{MPIM}
\end{equation}
where $h_{\rm pl}=H_{\rm g,pl}/r_{\rm pl}$ is the aspect ratio of the disc at the orbital position of the planet. We define $r_{\rm PIM}$ as the orbital position where the planet mass $M_{\rm pl}$ (we fix it as $20M_{\rm E}$) is equal to $M_{\rm PIM}$. The planet crosses the orbital position of $r=r_{\rm PIM}$ during its inward migration outside the snowline when $\alpha\leq10^{-2.6}$ (see Section \ref{regions} for details).

The ratio of the pressure gradient to the gravity of the central star, $\eta$, is important, because it determines the direction and speed of the drift of the particles (see Eq. (\ref{vdrift})). We calculate the ratio as
\begin{equation}
\eta=-\dfrac{1}{2}\left(\dfrac{H_{\mathrm{g}}}{r}\right)^{2}\dfrac{\partial \ln{\rho_{\mathrm{g}}c_{\mathrm{s}}^{2}}}{\partial \ln{r}},
\label{eta}
\end{equation}
where $\rho_{\rm g}=\Sigma_{\rm g}/(\sqrt{2\pi}H_{\rm g})$ is the (local) gas density in the midplane. We here define $r_{\eta0}$ as the orbital position where $\eta$ is zero (due to the cavity of the gas disc by the planet) when the planet is at $r=r_{\rm PIM}$. 

We consider the Type I migration of the planet. The migration timescale depends on the planet mass and the structures of the gas and temperature of the disc \citep{tan02,ida08a},
\begin{equation}
\tau_{\rm mig}=\dfrac{1}{2.728+1.082p}\left(\dfrac{c_{\rm s,pl}}{r_{\rm pl}\Omega_{\rm K,pl}}\right)^{2}\dfrac{M_{*}}{M_{\rm pl}}\dfrac{M_{*}}{r_{\rm pl}^{2}\Sigma_{\rm g,unp}}\Omega_{\rm K,pl}^{-1}.
\label{tmig}
\end{equation}
We assume the planet is at $r=50~{\rm au}$ at $t=0$ and migrates inward with a velocity equal to $v_{\rm pl}=-r_{\rm pl}/\tau_{\rm mig}$. We consider the reduction of the migration speed and the shift from the Type I to Type II migration due to the deep gap formation with the planetary growth by gas accretion in Section \ref{realistic}. We also investigate the cases where the planet forms later in that section.

\subsection{Dust evolution} \label{dustevolution}
We include in our model the evolution of dust particles in the gas disc model. We use a single-sized dust evolution model proposed by \citet{sat16}, which assumes that $m_{\rm d}$ singly peaks the mass distribution of dust at each orbit $r$. We calculate the radial distribution of the surface density of dust particles, $\Sigma_{\rm d}$, and their peak mass, $m_{\rm d}$, by solving the following equations, Eqs. (\ref{continuity}) and (\ref{growth}), simultaneously. The subscript ``d'' indicates the properties of the dust particles.

We consider the evolution of compact and spherical dust particles, the mass of a single particle being $m_{\rm d}=(4\pi/3)R_{\rm d}^{3}\rho_{\rm int}$, where $R_{\rm d}$ is the radius of the particles, and $\rho_{\rm int}=1.4$ and $3.0~{\rm g~cm^{-3}}$ are the internal density of the icy and rocky particles, respectively. Here, we assume that the particles are icy and rocky outside and inside the snowline, respectively.

The continuity equation of the dust particles is,
\begin{equation}
\dfrac{\partial\Sigma_{\rm d}}{\partial r}+\dfrac{1}{r}\dfrac{\partial}{\partial r}\left(rv_{r}\Sigma_{\rm d}-\dfrac{\nu}{1+{\rm St}^{2}}r\Sigma_{\rm g}\dfrac{\partial Z_{\Sigma}}{\partial r}\right)=0,
\label{continuity}
\end{equation}
where $\Sigma_{\rm d}$ is the dust surface density, $v_{\rm r}$ is the radial velocity of dust, $\nu=\alpha c_{\rm s}H_{\rm g}$ is the gas viscosity, and $Z_{\rm \Sigma}=\Sigma_{\rm d}/\Sigma_{\rm g}$ is the dust-to-gas surface density ratio. The Stokes number (stopping time normalized by Kepler time), ${\rm St}=t_{\rm stop}\Omega_{\rm K}$, determines the motion of the particles. The first and second terms in the parentheses represent the drift and diffusion of the particles, respectively.

The Stokes number of the dust particles is
\begin{equation}
{\rm St}=\dfrac{\pi}{2}\dfrac{\rho_{\rm int}R_{\rm d}}{\Sigma_{\rm g}}\max\left(1,\dfrac{4R_{\rm d}}{9\lambda_{\rm mfp}}\right),
\label{Stokes}
\end{equation}
where $\lambda_{\rm mfp}=m_{\rm g}/(\sigma_{\rm mol}\rho_{\rm g})$ is the mean free path of the gas molecules. Their collisional cross section is $\sigma_{\rm mol}=2\times10^{-15}{\rm cm}^{2}$.

The radial drift velocity of the particles is calculated by \citep{whi72,ada76,wei77}
\begin{equation}
v_{\rm drift}=-2\dfrac{\rm St}{\rm St^{2}+1}\eta v_{\rm k},
\label{vdrift}
\end{equation}
where $v_{\rm k}=r\Omega_{\rm k}$ is the Kepler velocity. The radial velocity of the particles due to their diffusion is
\begin{equation}
v_{\rm diff}=-\dfrac{\nu}{1+{\rm St}^{2}}\dfrac{1}{r}\dfrac{\partial\ln{Z_{\rm\Sigma}}}{\partial\ln{r}}.
\label{vdiff}
\end{equation}
The total radial velocity of the particles is $v_{\rm r}=v_{\rm drift}+v_{\rm diff}$. We reduce the inward dust mass flux, $\dot{M}_{\rm d}=-2\pi rv_{\rm r}\Sigma_{\rm d}$, just inside the snowline to be half of that just outside when $v_{\rm r}<0$ and increase the flux outside the snowline to the double of that just inside when $v_{\rm r}>0$ to express the evaporation and re-condensation of the icy material of the particles.

The growth of the particles due to their mutual collision is \citep{sat16}
\begin{equation}
\dfrac{\partial m_{\rm d}}{\partial t}+v_{\rm r}\dfrac{\partial m_{\rm d}}{\partial r}=\epsilon_{\rm grow}\dfrac{2\sqrt{\pi}R_{\rm d}^{2}\Delta v_{\rm dd}}{H_{\rm d}}\Sigma_{\rm d},
\label{growth}
\end{equation}
where $\epsilon_{\rm grow}$, $\Delta v_{\rm dd}$, and $H_{\rm d}$ are the sticking efficiency, collision velocity, and dust scale height, respectively.

The particles break up rather than merge when the collision speed is too fast. We model the sticking efficiency as \citep{oku16}
\begin{equation}
\epsilon_{\rm grow}=\min\left\{1, -\dfrac{\ln{(\Delta v_{\rm dd}/v_{\rm cr})}}{\ln{5}}\right\},
\label{stfrag}
\end{equation}
where the critical fragmentation velocity for the collision of rocky and icy particles are $v_{\rm cr}=1$ and $10~{\rm m~s^{-1}}$, respectively \citep[e.g.][]{blu00,wad09}\footnote{This expression can be used even if dust grows to meter-size, because numerical simulations show that the fragmentation dose not depend on the number of monomers of dust aggregates \citep{wad09}.}.

The collision velocity between the dust particles is
\begin{equation}
\Delta v_{\rm dd}=\sqrt{\Delta v_{\rm B}^{2}+\Delta v_{\rm drift}^{2}+\Delta v_{\rm \phi}^{2}+\Delta v_{\rm z}^{2}+\Delta v_{\rm diff}^{2}},
\label{vdd}
\end{equation}
where $\Delta v_{\rm B}$, $\Delta v_{\rm drift}$, $\Delta v_{\rm \phi}$, $\Delta v_{\rm z}$, and $\Delta v_{\rm diff}$ are the relative velocities induced by their Brownian motion, radial drift, azimuthal drift, vertical sedimentation, and diffusion, respectively \citep{oku12}. The relative velocity induced by Brownian-motion between the particles with the same mass is $\Delta v_{\rm B}=\sqrt{16k_{\rm B}T/(\pi m_{\rm d})}$. The relative velocity induced by the radial drift is $\Delta v_{\rm drift}=|v_{\rm drift}({\rm St}_{1})-v_{\rm drift}({\rm St}_{2})|$, where ${\rm St}_{1}$ and ${\rm St}_{2}$ are the Stokes numbers of the two colliding particles. The relative velocity induced by the azimuthal drift is $\Delta v_{\rm \phi}=|v_{\rm \phi}({\rm St}_{1})-v_{\rm \phi}({\rm St}_{2})|$, where $v_{\rm \phi}=-\eta v_{\rm K}/(1+{\rm St}^{2})$, and that by the vertical motion is $\Delta v_{\rm z}=|v_{\rm z}({\rm St}_{1})-v_{\rm z}({\rm St}_{2})|$, where $v_{\rm z}=-\Omega_{\rm K}{\rm St}H_{\rm d}/(1+{\rm St})$. We assume ${\rm St}_{2}=0.5{\rm St}_{1}$, because the single-size simulation reproduces the results by a full-size simulation very well with that assumption \citep{sat16}. For the relative velocity induced by diffusion, we use following three limiting expressions derived from \citep{orm07},
\begin{equation}
\Delta \varv_{\rm diff}=
\begin{cases}
\sqrt{\alpha}c_{\rm s}{\rm Re}_{\rm t}^{1/4}\left|{\rm St_{1}}-{\rm St_{2}}\right|, & {\rm St_{1}}\ll {\rm Re}_{\rm t}^{-1/2}, \\
\sqrt{3\alpha}c_{\rm s}{\rm St}_{1}^{1/2}, & {\rm Re}_{\rm t}^{-1/2} \ll {\rm St_{1}}\ll 1, \\
\sqrt{\alpha}c_{\rm s}\left(\dfrac{1}{1+{\rm St_{1}}}+\dfrac{1}{1+{\rm St_{2}}}\right)^{1/2}, & 1\ll {\rm St_{1}}.
\end{cases}
\label{St}
\end{equation}
where ${\rm Re_{t}}=\nu/\nu_{\rm mol}$ is the turbulence Reynolds number. The molecular viscosity is $\nu_{\rm mol}=v_{\rm th}\lambda_{\rm mfp}/2$, where $v_{\rm th}=\sqrt{8/\pi}c_{\rm s}$ is the thermal gas velocity.

The dust scale height is given by \citep{you07},
\begin{equation}
H_{\rm d}=H_{\rm g}\left(1+\dfrac{{\rm St}}{\alpha}\dfrac{1+2{\rm St}}{1+{\rm St}}\right)^{-1/2},
\label{Hd}
\end{equation}
and the midplane dust density is $\rho_{\rm d,mid}=\Sigma_{\rm d}/(\sqrt{2\pi}H_{\rm d})$.

\subsection{Planetesimal formation} \label{planetesimal}
We calculate when, where, and how much planetesimals form in our scenario. In this work, we consider two mechanisms of planetesimals formation: streaming instability and mutual collision of particles. First, we consider the condition for planetesimal formation by streaming instability. Streaming instability can enhance the accumulation of dust particles, which helps the condition for gravitational instability, $\rho_{\rm d, mid}>\rho_{\rm Roche}\equiv9M_{*}\Omega_{\rm K}^{2}/(4\pi G)$, be reached. We define the condition for planetesimal formation as the dust-to-gas density ratio on the midplane, $Z_{\rho}\equiv\rho_{\rm d, mid}/\rho_{\rm d, gas}$, is larger than the critical density ratio $\epsilon_{\rm crit}=1$ \citep{you05,joh07,dra14}. We also consider the case in which $\epsilon_{\rm crit}$ depends on ${\rm St}$ in Section \ref{St-dependence}. We assume that planetesimal formation only occurs outside the orbit of the migrating planet in order to focus on the planetesimal formation at the gas pressure bump created by the planet\footnote{Without this assumption, planetesimals episodically form inside the planetary orbit due to waves forming in radial profiles of dust when the pebble front crosses the gap, which should not be real.}.

The change of the planetesimal surface density due to streaming instability is
\begin{equation}
\dfrac{{\rm d}\Sigma_{\rm{pls,SI}}}{{\rm d}t}=x_{\rm SI}\Sigma_{\rm{d}}=\dfrac{\epsilon_{\rm SI}}{\tau_{\rm SI}}\Sigma_{\rm d},
\label{dSigmaSI}
\end{equation}
where the efficiency of streaming instability is assumed $\epsilon_{\rm SI}=0.1$ \citep{sch18}\footnote{Although $\epsilon_{\rm SI}$ has been treated as a free parameter in previous works \citep[e.g.][]{dra14}, its variety is implicitly expressed together with the variety of $\tau_{\rm SI}$ in this work.}. The timescale of streaming instability, $\tau_{\rm SI}$, is an important parameter of this work. We consider the cases with short timescale ($\tau_{\rm SI}=10~{\rm years}$ \citep{you05,joh07}, in Section \ref{short-tSI}) and with long timescale ($\tau_{\rm SI}=10^{3}T_{\rm K}$, where $T_{\rm K}=2\pi/\Omega_{\rm K}$ is the orbital period \citep{dra16}, in Section \ref{long-tSI}). We also investigate the cases where the timescale depends on ${\rm St}$ in Section \ref{St-dependence}.

We also consider planetesimal formation due to mutual collision of particles. When the dust radius $R_{\rm d}$ is larger than $R_{\rm d,max}=1~{\rm m}$, we define that the particles become planetesimals. This is a valid definition, because the rapid growth of particles starts when the particles are smaller than $1~{\rm m}$ (see the third column of Fig. \ref{fig:evolution-dust-longtSI})\footnote{We also check that the particle radius becomes much larger than $1~{\rm m}$ immediately when we do not consider the planetesimal formation due to mutual collision.}. In every time step, we check this condition after checking the condition for streaming instability.

Although we use a single-sized dust evolution model in this work, the particles have size frequency distribution (SFD) at each $r$ in reality. Hence, we assume an ``imaginary'' SFD and regard the mass of the particles larger than $R_{\rm d,max}$ in the SFD as the mass of newly formed planetesimals due to mutual collision. We assume the SFD as $dN\propto a^{-q_{\rm d}}da$, where $N$ is the number of the particles larger than $a$, and the minimum radius as $R_{\rm d,min}$. In that case, the change of the planetesimal surface density due to mutual collision is
\begin{equation}
\dfrac{{\rm d}\Sigma_{\rm{pls,MC}}}{{\rm d}t}=\dfrac{\rm d}{{\rm d}t}\left(\dfrac{R_{\rm d}^{4-q_{\rm d}}-R_{\rm d,max}^{4-q_{\rm d}}}{R_{\rm d}^{4-q_{\rm d}}-R_{\rm d,min}^{4-q_{\rm d}}}\Sigma_{\rm d}\right).
\label{dSigmaMC}
\end{equation}
Here, we assume that $R_{\rm d,min}=0.1~{\rm \mu m}$ and $q_{\rm d}=3.5$.

The total planetesimal formation rate is
\begin{equation}
\dfrac{{\rm d}\Sigma_{\rm{pls,tot}}}{{\rm d}t}=\dfrac{{\rm d}\Sigma_{\rm{pls,SI}}}{{\rm d}t}+\dfrac{{\rm d}\Sigma_{\rm{pls,MC}}}{{\rm d}t},
\label{dSigmatot}
\end{equation}
where the total planetesimal surface density is $\Sigma_{\rm{pls,tot}}=\Sigma_{\rm{pls,SI}}+\Sigma_{\rm{pls,MC}}$. At the same time, the dust surface density is reduced by planetesimal formation,
\begin{equation}
\dfrac{{\rm d}\Sigma_{\rm d}}{{\rm d}t}=-\dfrac{{\rm d}\Sigma_{\rm{pls,tot}}}{{\rm d}t}.
\label{dSigmad}
\end{equation}

\section{Results} \label{results}
\subsection{Short streaming instability timescale} \label{short-tSI}
\subsubsection{Evolution of gas and solids} \label{evolution}
\begin{figure}[htbp]
\centering
\includegraphics[width=0.65\linewidth]{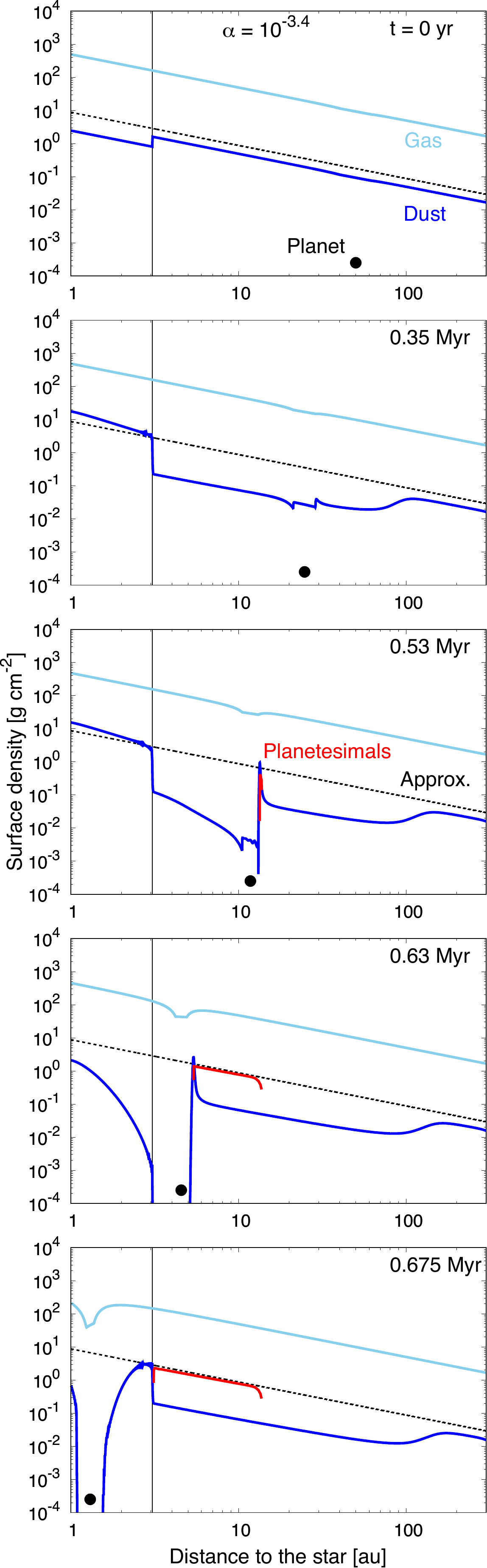}
\caption{Time evolution of the surface density of gas, dust, and formed planetesimals, with $\alpha=10^{-3.4}$. The sky blue, blue, and red curves represent the profiles of gas, dust, and planetesimals, respectively. All planetesimals form by streaming instability. The black dashed lines represent the approximation of the planetesimal surface density by Eq. (\ref{Sigmapl_est}) with $\dot{M}_{\rm d,pl}=1.5\times10^{-4}~M_{\rm E}~{\rm yr}^{-1}$. The circles and black vertical lines represent the orbital positions of the planet and the snowline, respectively.}
\label{fig:evolution-Sigma}
\end{figure}

We show in Fig. \ref{fig:evolution-Sigma} the results with the short SI (streaming instability) timescale ($\tau_{\rm SI}=10~{\rm yrs}$). The figure represents the evolution of the surface densities of the gas, dust, and planetesimals with $\alpha=10^{-3.4}=4\times10^{-4}$. The embedded planet carves a deeper gap in the gas, as it migrates inward (the sky blue curves). Thus, the dust particles accumulate more at the outer edge of the gap as the planet migrates inward (blue curves). The figure also shows that the pebble front, the orbital position where the drift timescale of dust becomes shorter than the growth timescale, and dust (pebbles) starts to drift toward the central star \citep{lam14b}, moves outward (the bold curves around $100~{\rm au}$). This phenomenon is not related to the embedded planet. Figure \ref{fig:evolution-Sigma} also shows that planetesimals form by streaming instability between the snowline and the orbital position where the inward drift of dust starts to be halted (red curves). The formed planetesimal surface density is about $2~{\rm g~cm^{-2}}$ at the inner edge and about $0.6~{\rm g~cm^{-2}}$ at the outer edge of the formation region.

\begin{figure*}[tbp]
\centering
\includegraphics[width=0.9\linewidth]{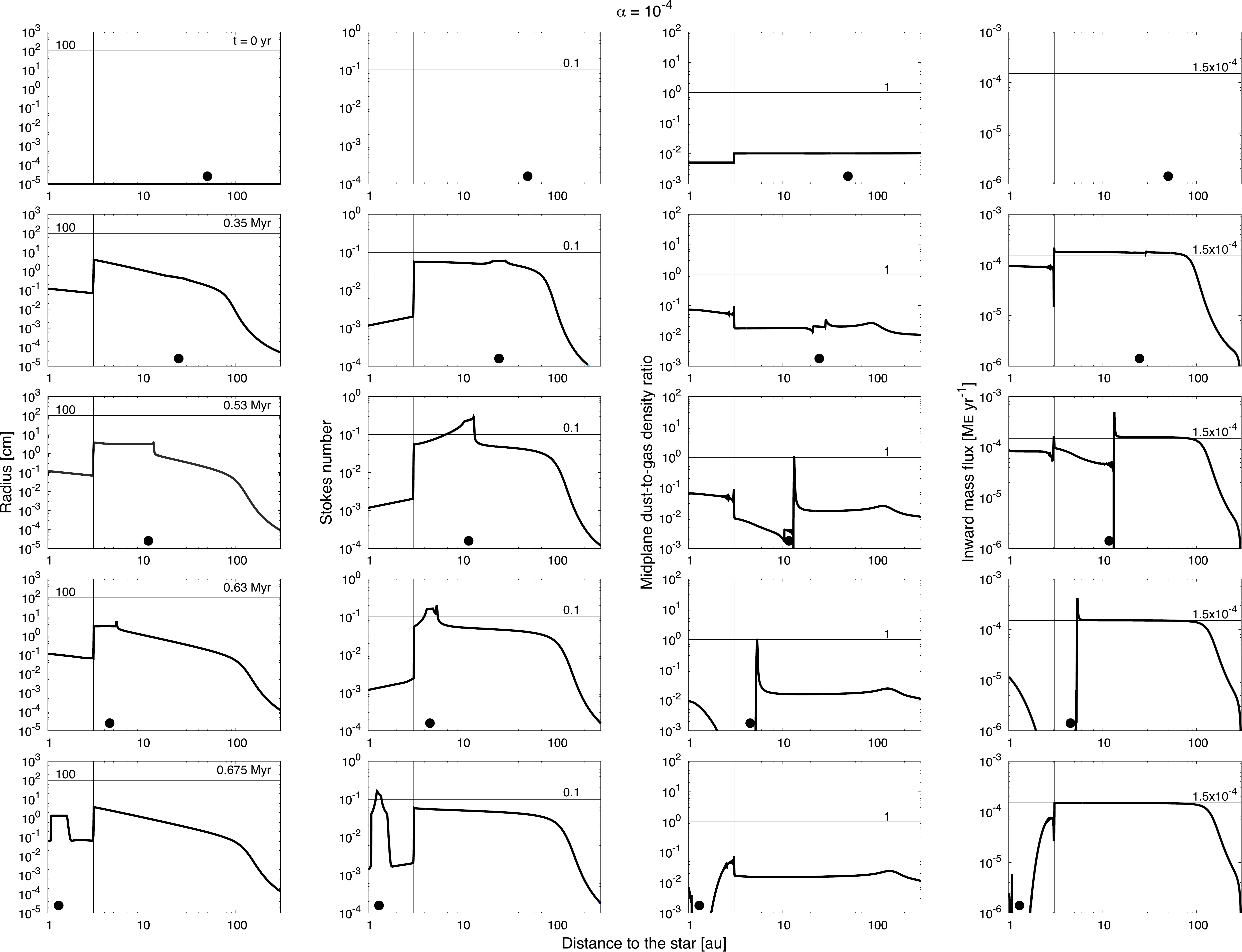}
\caption{Time evolution of the detailed profiles of dust with $\alpha=10^{-3.4}$. The first to fourth columns from the left represent the radial profiles of the radius, Stokes number, midplane dust-to-gas density ratio, and inward mass flux, respectively. The first to fifth rows from the top represent the profiles at $t=0, 0.35, 0.53, 0.63$ and $0.75~{\rm Myr}$, respectively. The horizontal lines represent the critical/important values of each profile. The vertical lines represent the position of the snowline. The circles represent the orbital positions of the planet.}
\label{fig:evolution-dust}
\end{figure*}

Figure \ref{fig:evolution-dust} shows the detailed evolution of the dust. The first column shows that the radius is smaller than $R_{\rm d,max}=1~{\rm m}$, so that the all planetesimals should be formed by streaming instability. The Stokes number of dust is smaller than $0.1$ when the dust drifts inward, but it increases when the dust is accumulated (the second column). The Stokes number in the accumulation is about $0.1$ when the accumulation starts ($t=0.35~{\rm Myr}$) and is larger than $0.1$ in the full accumulation ($t=0.53$ and $0.63~{\rm Myr}$). The accumulation of dust is not enough for the condition for the planetesimal formation ($Z_{\rho}\geq1$) in the beginning of the accumulation ($t=0.35~{\rm Myr}$), but it reaches the condition when the drift of dust is stopped ($t=0.53~{\rm Myr}$) (the third column). After that, the particles disappear inside the outer edge of the gap, and the inward mass flux of the drifting dust is zero (the fourth column). This condition for the rapid accumulation of dust is consistent with the one proposed by \citet{tak21}, ${\partial\dot{M}_{\rm d}}/{\partial r}<0$ (Eq. (30) in the paper). The inward mass flux is uniform, almost constant, and equal to $\dot{M}_{\rm d,pl}=1.5\times10^{-4}~M_{\rm E}~{\rm yr}^{-1}$ outside the gap.

Once the planet (and the gas pressure bump) crosses the snowline, the midplane dust-to-gas density ratio decreases and planetesimals cannot form. This is because the Stokes number of the dust becomes smaller due to the fragile rocky particles (the second column), which makes the vertical diffusion of dust more efficiently and $\rho_{\rm d,mid}$ lower (the third column). The dust mass flux is about half of the one outside the snowline (the fourth column), which is another reason.
 
\begin{figure}[tbp]
\centering
\includegraphics[width=0.9\linewidth]{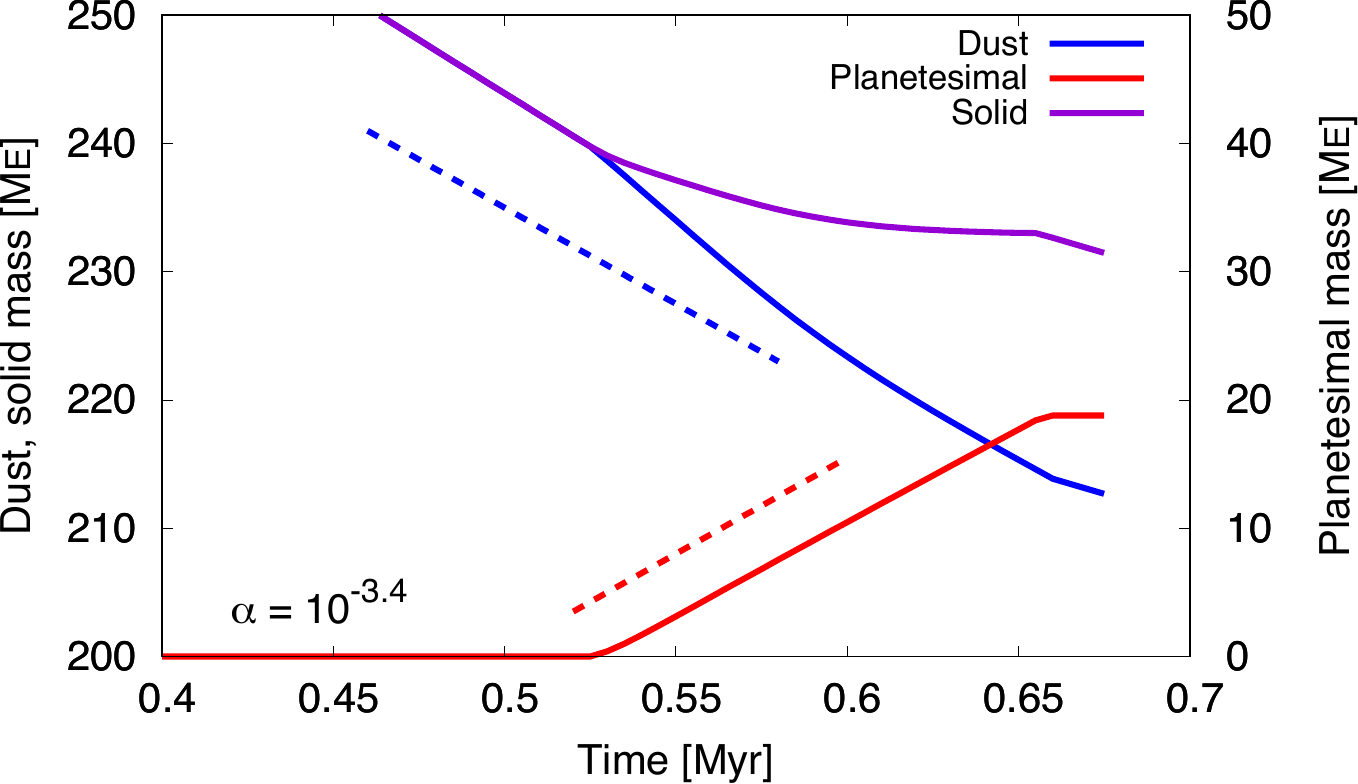}
\caption{Time evolution of the dust, planetesimal, and solid (sum of the dust and planetesimals) mass with $\tau_{\rm SI}=10~{\rm yr}$ and $\alpha=10^{-3.4}$. The blue, red, and purple curves represent the profiles of the dust, planetesimal, and solid, respectively. The dashed blue line represents the slope of the dust mass assuming constant loss with $\dot{M}_{\rm d}=1.5\times10^{-4}~M_{\rm E}~{\rm yr}^{-1}$. The red dashed line represents the slope of the cumulative mass of the dust drifting into the planetesimal formation place with the same mass flux.}
\label{fig:evolution-mass-shorttSI}
\end{figure}

We proposed an approximate expression of the planetesimal surface density in Paper 1,
\begin{eqnarray}
\Sigma_{\rm{pls,est}}&\equiv&\dfrac{\dot{M}_{\rm d}}{2\pi rv_{\rm pl}} \nonumber \\
&=&8.8\left(\dfrac{\Sigma_{\rm g,1au}}{500~{\rm g~cm}^{-2}}\right)^{-1}\left(\dfrac{T_{\rm 1au}}{280~{\rm K}}\right)\left(\dfrac{M_{\rm pl}}{20~M_{\rm E}}\right)^{-1} \nonumber \\
&\times&\left(\dfrac{M_{*}}{M_{\odot}}\right)^{1/2}\left(\dfrac{\dot{M}_{\rm d}}{1.5\times10^{-4}~M_{\rm E}~{\rm yr}^{-1}}\right)\left(\dfrac{r}{\rm au}\right)^{-1}{\rm g~cm}^{-2},
\label{Sigmapl_est}
\end{eqnarray}
where $\dot{M}_{\rm d}$ is the inward dust mass flux (see Appendix \ref{analytical} for more general expressions). Figure \ref{fig:evolution-Sigma} shows that our results are very well approximated when we substitute $\dot{M}_{\rm d}=1.5\times10^{-4}~M_{\rm E}~{\rm yr}^{-1}$, which is obtained from our results (see the fourth column of Fig. \ref{fig:evolution-dust}), into Eq. (\ref{Sigmapl_est}). This means that all dust drifting into the formation place (around where $\eta=0$) is converted immediately to planetesimals once the formation starts. This is also shown in Fig. \ref{fig:evolution-mass-shorttSI} that the planetesimal mass (red solid curve) increases linearly along with the slope of the cumulative dust mass drifting into the formation place with $\dot{M}_{\rm d,pl}=1.5\times10^{-4}~M_{\rm E}~{\rm yr}^{-1}$ (red dashed line). The dust mass (blue solid curve) also decreases linearly before the beginning of planetesimal formation along with the slope of the dust mass assuming constant loss with the same mass flux with $\dot{M}_{\rm d,pl}$ (blue dashed line), but it decreases more once the planetesimals start to form ($0.53\leq t\leq0.6~{\rm Myr}$). This is because, although the rate of the mass converting from dust to planetesimals is the same with the one losing from the disc before the planetesimal formation starts, the dust exists inside the gas pressure bump continues to disappear gradually also after the formation starts (see Fig. \ref{fig:evolution-Sigma}). This is also the reason why the slope of the solid (sum of the dust and planetesimals) mass profile in Fig. \ref{fig:evolution-mass-shorttSI} gradually becomes zero. Once the gas pressure bump crosses the snowline, the planetesimal formation stops, and the increase of the planetesimal mass also stops ($t=0.66~{\rm Myr}$).

\subsubsection{Planetesimal formation regions} \label{regions}
\begin{figure}[tbp]
\centering
\includegraphics[width=0.9\linewidth]{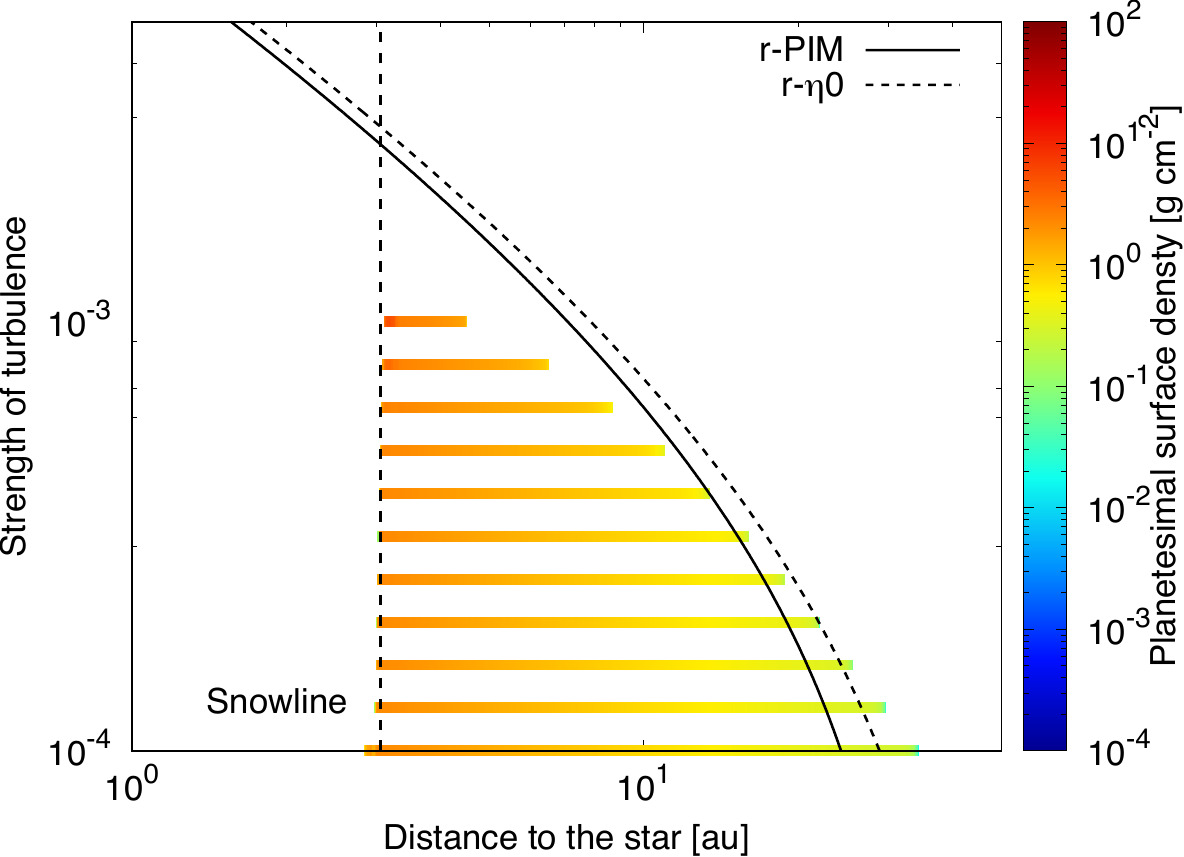}
\caption{Planetesimal formation regions with $\tau_{\rm SI}=10~{\rm yr}$ and various $\alpha$. The colour represents the planetesimal surface density. The solid and dotted curves represent $r_{\rm PIM}$ and $r_{\eta0}$, respectively. The vertical dashed line is the snowline.}
\label{fig:regions-shorttSI}
\end{figure}

We then investigate the formation regions of planetesimals by changing the value of $\alpha$. Figure \ref{fig:regions-shorttSI} shows that planetesimals form when $10^{-4}\leq\alpha\leq10^{3}$, which is broadly consistent with the measured value of $\alpha$ in a lot of observed protoplanetary discs \citep{pin22}. The figure also shows the mechanism of the planetesimal formation is streaming instability in all cases. This is because, the dust being piled up at the bump convert to planetesimals with the instability before they grow to planetesimals by mutual collision. We find that belt-like planetesimal formation regions exist between the snowline and the position where the planet reaches its pebble-isolation mass (Eq. (\ref{MPIM})), $r_{\rm PIM}$. Planetesimals do not form inside the snowline as we explained in Section \ref{evolution}. The pebble-isolation mass is the mass the planet needs in order to make the gap deep enough to stop the dust (pebble) accretion, meaning that all of the dust drifting into the region piles up, which in turn triggers streaming instability. For the calculation of $r_{\rm PIM}$ in this work, we fix the Stokes number as ${\rm St}=0.1$.

\begin{figure}[tbp]
\centering
\includegraphics[width=0.9\linewidth]{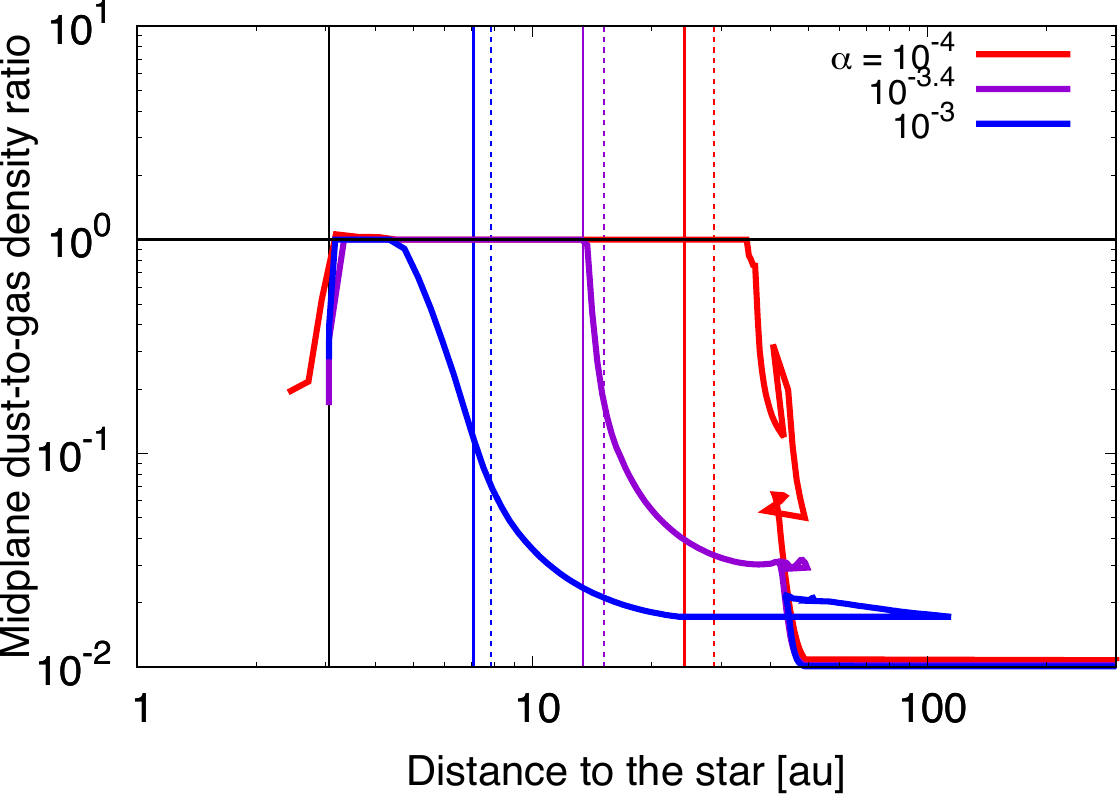}
\caption{Trajectory of the largest $Z_{\rho}$ outside the orbit of the planet with $\tau_{\rm SI}=10~{\rm yr}$. The red, purple, and blue curves represent the cases with $\alpha=10^{-4}$, $10^{-3.4}$, and $10^{-3}$, respectively. The solid and dotted vertical lines represent $r_{\rm PIM}$ and $r_{\eta0}$ with each $\alpha$, respectively.}
\label{fig:rhodgmax}
\end{figure}

When $\alpha<10^{-3.4}$, the outer edge of the formation region is slightly outside $r_{\rm PIM}$, and the distance between the two orbital positions is larger as $\alpha$ is smaller. On the other hand, when $\alpha>10^{-3.4}$, the outer edge is inner than $r_{\rm PIM}$, and the distance between the two orbital positions is larger as $\alpha$ is larger. This is because to get the condition for planetesimal formation, $Z_{\rho}$ needs to increase beyond unity against the turbulence. In other words, it is the diffusion of the particles, which prevents the accumulation of the dust. Figure \ref{fig:rhodgmax} shows that the orbital position where the largest $Z_{\rho}$ outside the planetary orbit reaches unity is outside $r=r_{\rm PIM}$ when $\alpha=10^{-4}$. The position is on $r=r_{\rm PIM}$ when $\alpha=10^{-3.4}$ and is inside when $\alpha=10^{-3}$. This result is consistent with Fig. \ref{fig:regions-shorttSI}. The reasons why the profiles in Fig. \ref{fig:rhodgmax} wander at their outer parts are that the pebble front has the largest value of $Z_{\rho}$ until the rapid accumulation of dust at the gas pressure bump starts, and the pebble front also makes waves in the dust profiles when it crosses the gap created by the planet (especially when $\alpha=10^{-4}$).

\begin{figure}[tbp]
\centering
\includegraphics[width=0.9\linewidth]{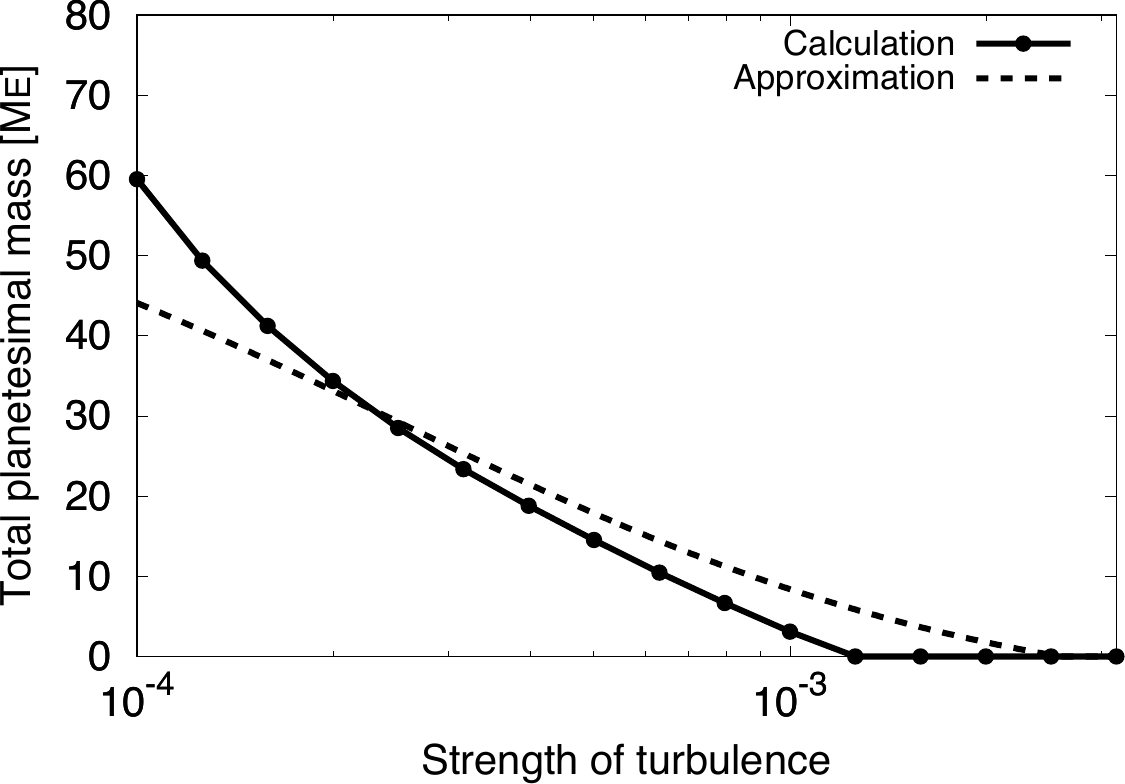}
\caption{Total mass of the formed planetesimals with $\tau_{\rm SI}=10~{\rm yr}$ and various $\alpha$. The solid and dashed curves represent the results of our calculations and the approximation by Eq. (\ref{Mplstotest}), respectively.}
\label{fig:planetesimal-mass}
\end{figure}

Figure \ref{fig:planetesimal-mass} shows that the total mass of the formed planetesimals is larger as $\alpha$ is smaller due to the $\alpha$ dependence of the outer edge of the formation region. When $\alpha=10^{-4}$, the total mass reaches about $60~M_{\rm E}$. We estimate the total mass of the planetesimals by
\begin{equation}
M_{\rm pls,tot,est}\equiv\int^{r_{\rm PIM}}_{r_{\rm SL}}2\pi r\Sigma_{\rm pls,est}dr.
\label{Mplstotest}
\end{equation}
The figure shows this estimate roughly reproduces the results of our calculations. The difference at the high and low $\alpha$ is because the precise location of the outer edge of the planetesimal formation region is different from $r_{\rm PIM}$, as we explained above.

\subsection{Long streaming instability timescale} \label{long-tSI}
We then investigate the planetesimal formation with the long SI timescale ($\tau_{\rm SI}=10^{3}T_{\rm K}$). In the case of the short SI timescale, all planetesimals form with streaming instability independent from the strength of the turbulence. On the other hand, in the case of the long SI timescale, the formation mechanism depends on the turbulence strength.

\begin{figure*}[tbp]
\centering
\includegraphics[width=0.9\linewidth]{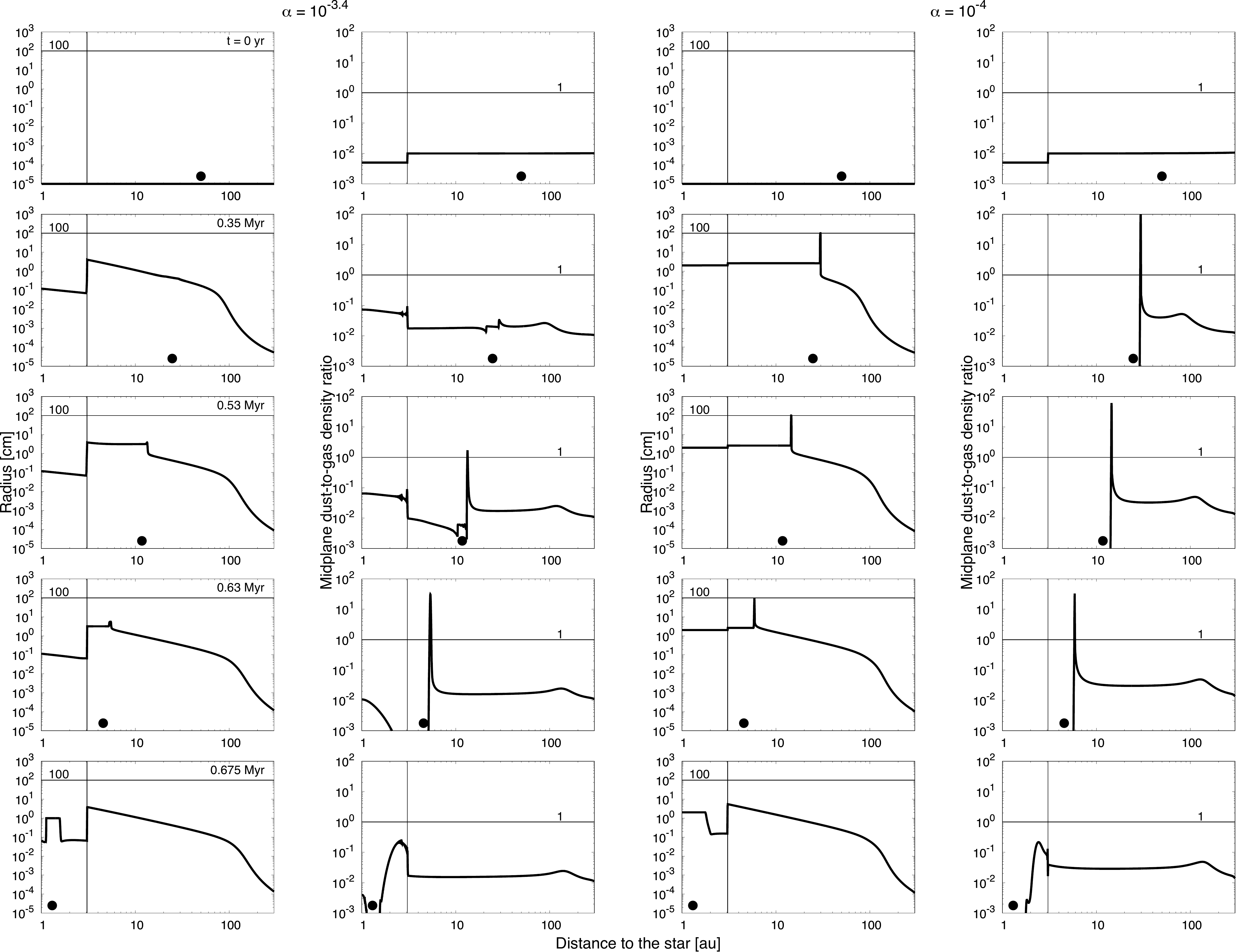}
\caption{Same as the first and third columns of Fig. \ref{fig:evolution-dust} but with $\tau_{\rm SI}=10^{3}T_{\rm K}$. The left two and right two columns represent the profiles with $\alpha=10^{-3.4}$ and $10^{-4}$, respectively.}
\label{fig:evolution-dust-longtSI}
\end{figure*}

\begin{figure*}
\centering
\includegraphics[width=0.45\linewidth]{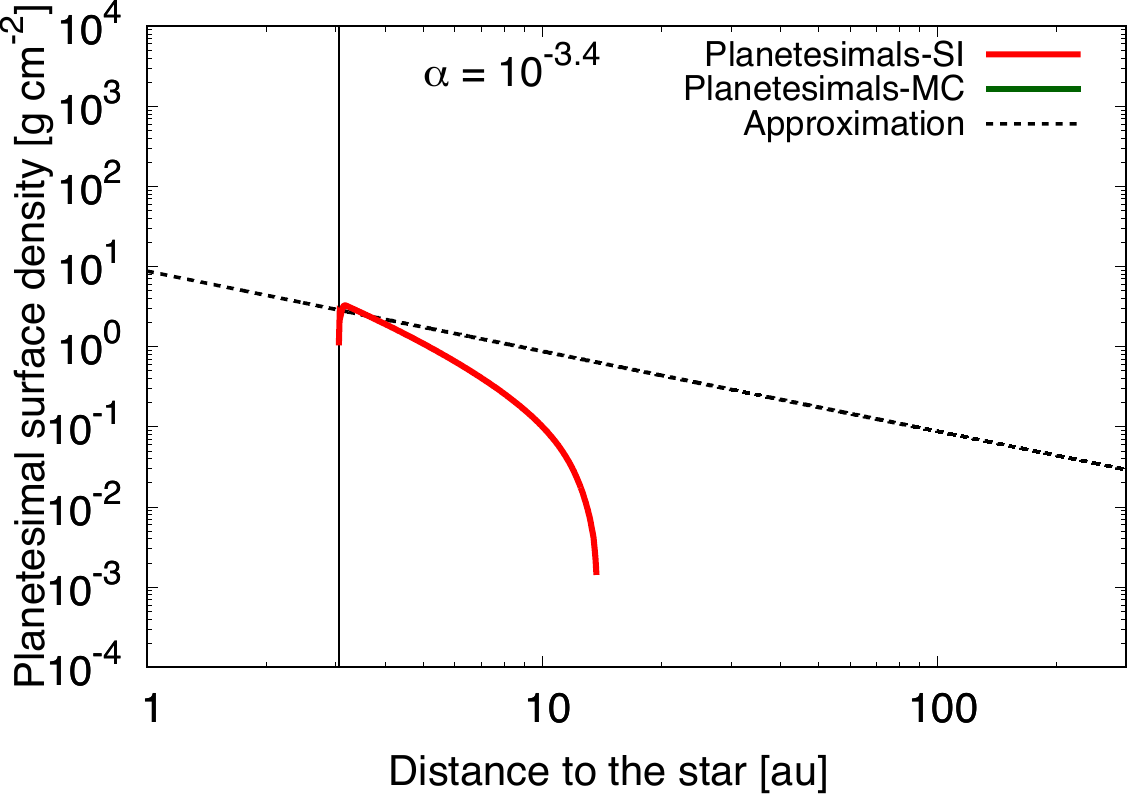}
\includegraphics[width=0.45\linewidth]{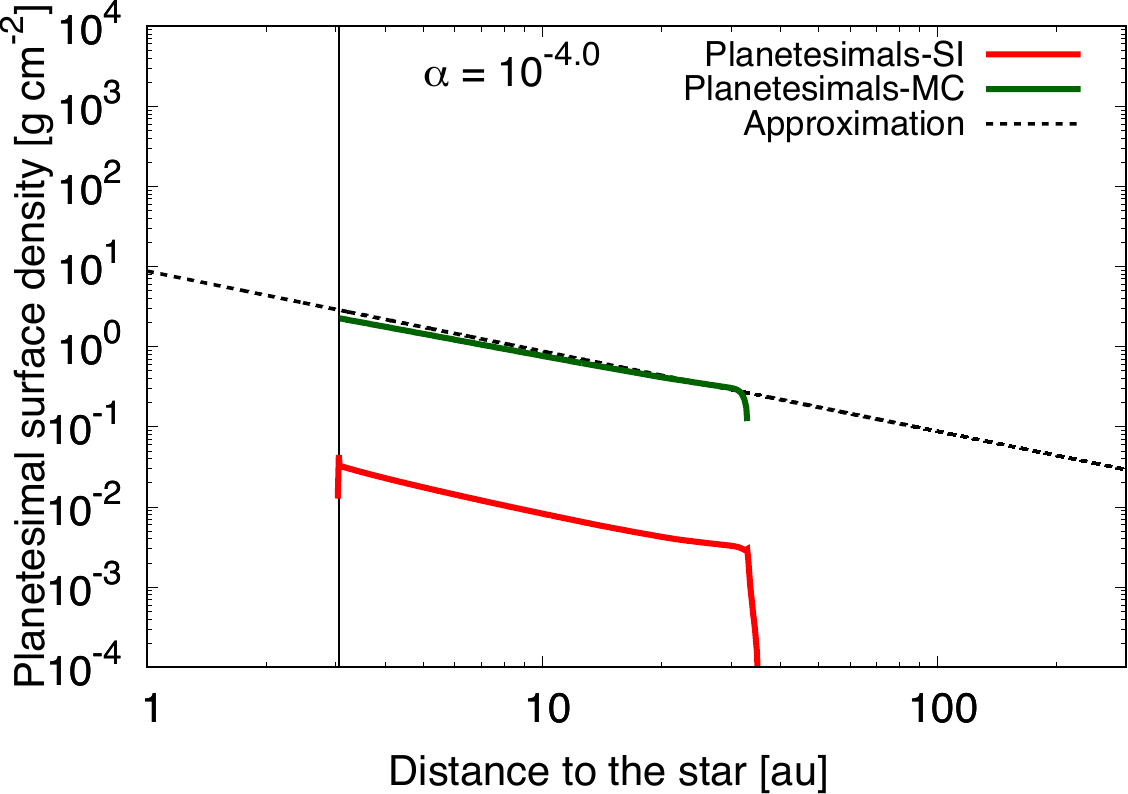}
\caption{Final profiles of the planetesimal surface density with $\tau_{\rm SI}=10^{3}T_{\rm K}$. The red and green curves represent the surface density of the planetesimals formed by streaming instability and mutual collision, respectively. The left and right panels represent the profiles with $\alpha=10^{-3.4}$ and $10^{-4}$, respectively.}
\label{fig:planetesimals-longtSI}
\end{figure*}

\begin{figure*}[tbp]
\centering
\includegraphics[width=0.45\linewidth]{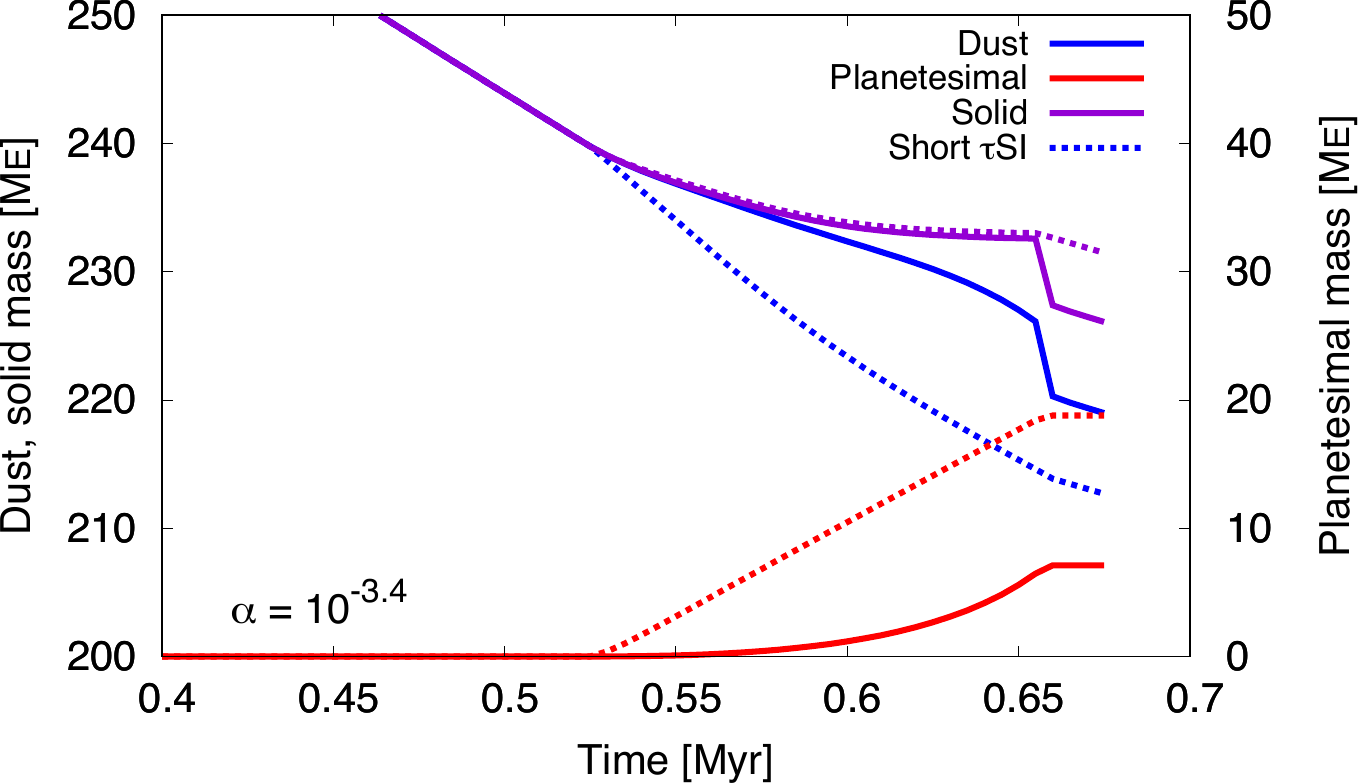}
\includegraphics[width=0.45\linewidth]{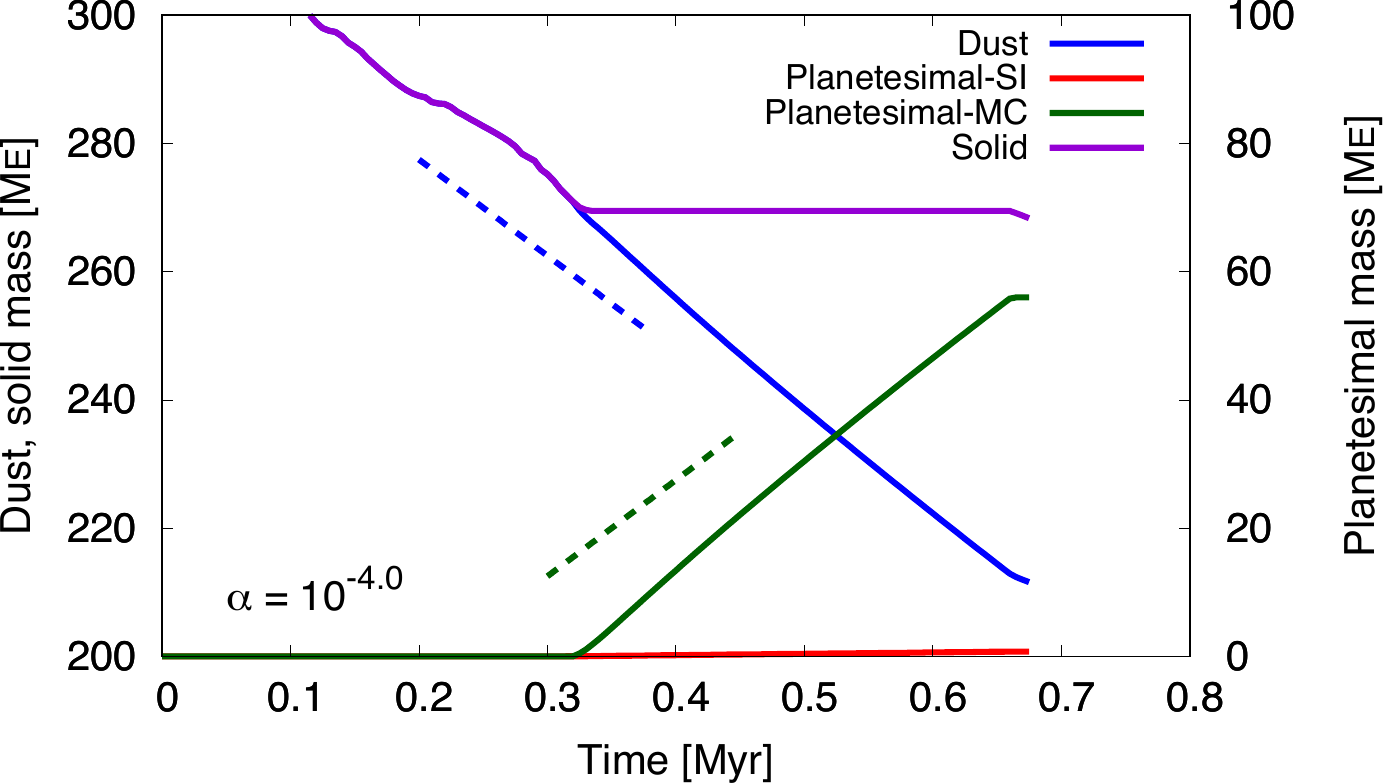}
\caption{Same as Fig. \ref{fig:evolution-mass-shorttSI} but with $\tau_{\rm SI}=10^{3}T_{\rm K}$. The left and right panels represent the profiles with $\alpha=10^{-3.4}$ and $10^{-4}$, respectively. In the left panel, the profiles with $\tau_{\rm SI}=10~{\rm yr}$ are also plotted as the dotted curves. All planetesimals form by streaming instability when $\alpha=10~{-3.4}$. In the right panel, the mass of the planetesimals formed by both streaming instability (red) and mutual collision (green) is plotted. The blue dashed line represents the slope of the dust mass assuming constant loss with $\dot{M}_{\rm d}=1.5\times10^{-4}~M_{\rm E}~{\rm yr}^{-1}$. The green dashed line represents the slope of the cumulative mass of the dust drifting into the planetesimal formation place with the same mass flux.}
\label{fig:evolution-mass-longtSI}
\end{figure*}

Figures \ref{fig:evolution-dust-longtSI} and \ref{fig:planetesimals-longtSI} represent the profiles of the dust evolution and the planetesimal surface density with the long SI timescale, respectively. The first column of Fig. \ref{fig:evolution-dust-longtSI} shows that the dust radius is smaller than $R_{\rm d,max}=1~{\rm m}$, so that the planetesimals are formed by streaming instability as in the case of short SI timescale. However, the left panel of Fig. \ref{fig:planetesimals-longtSI} shows that the radial profile of the planetesimal surface density is lower than the approximation (Eq. (\ref{Sigmapl_est})) in the outer part of the planetesimal formation region. This means that only parts of the drifting dust entering the formation place of planetesimals convert to planetesimals, because the approximation (Eq. (\ref{Sigmapl_est})) assumes that all dust converts to planetesimals immediately. The rest of the dust piles up there and makes $Z_{\rho}$ larger than $\epsilon_{\rm crit}=1$, the local condition for planetesimal formation (the second column of Fig. \ref{fig:evolution-dust-longtSI}). These interpretations are consistent with the time evolution of the dust and planetesimal mass shown in the left panel of Fig. \ref{fig:evolution-mass-longtSI}. The panel shows that the slope of the planetesimal mass with the long SI timescale (red solid curve) is much smaller than that with the short SI timescale (dotted red curve) (i.e., the case that all mass of the dust converting to the planetesimals) especially at the begging of the planetesimal formation ($0.55<t<0.6~{\rm Myr}$). The profiles of the solid mass (solid and dotted purple curves) are the same in both SI timescale, and the dust mass, assuming long SI timescale, does not decreases like the case we assume short SI timescale. This also means that dust particles not converted to the planetesimals pile up at the gas pressure bump. The sharp decrease of the dust (and solid) mass at $t=0.65~{\rm Myr}$ is because the piled-up dust evaporates when they cross the snowline.

On the other hand, the third column of Fig. \ref{fig:evolution-dust-longtSI} shows that the radius reaches $R_{\rm d,max}=1~{\rm m}$ when $\alpha=10^{-4}$. At the same time, the density ratio $Z_{\rho}$ is larger than $\epsilon_{\rm crit}=1$ (the fourth column). This means that the planetesimals are formed by both streaming instability and mutual collision. The right panel of Fig. \ref{fig:planetesimals-longtSI} shows that planetesimals are formed by both mechanisms but mainly by mutual collision when the turbulence is weak. The surface density of planetesimals formed by mutual collision is about $100$ times larger than the one of planetesimals formed by streaming instability. The panel also shows that the surface density of planetesimals formed by mutual collision is very well approximated by Eq. (\ref{Sigmapl_est}). However, dust also piles up at the formation place with $Z_{\rho}$ larger than $\epsilon_{\rm crit}=1$ (the fourth column of Fig.\ref{fig:evolution-dust-longtSI}), because $Z_{\rho}$ becomes easily large with $\alpha=10^{-4}$ (i.e., weak diffusion) compared to $\alpha=10^{-3.4}$. The right panel of Fig. \ref{fig:evolution-mass-longtSI} shows that all mass of dust drifting into the formation place converts to the mass of planetesimals (mainly) by mutual collision once the planetesimals start to form ($t=0.3~{\rm Myr}$). As a result, the solid mass (i.e., total mass of the dust and planetesimals) is conserved after that.

\begin{figure*}
\centering
\includegraphics[width=0.45\linewidth]{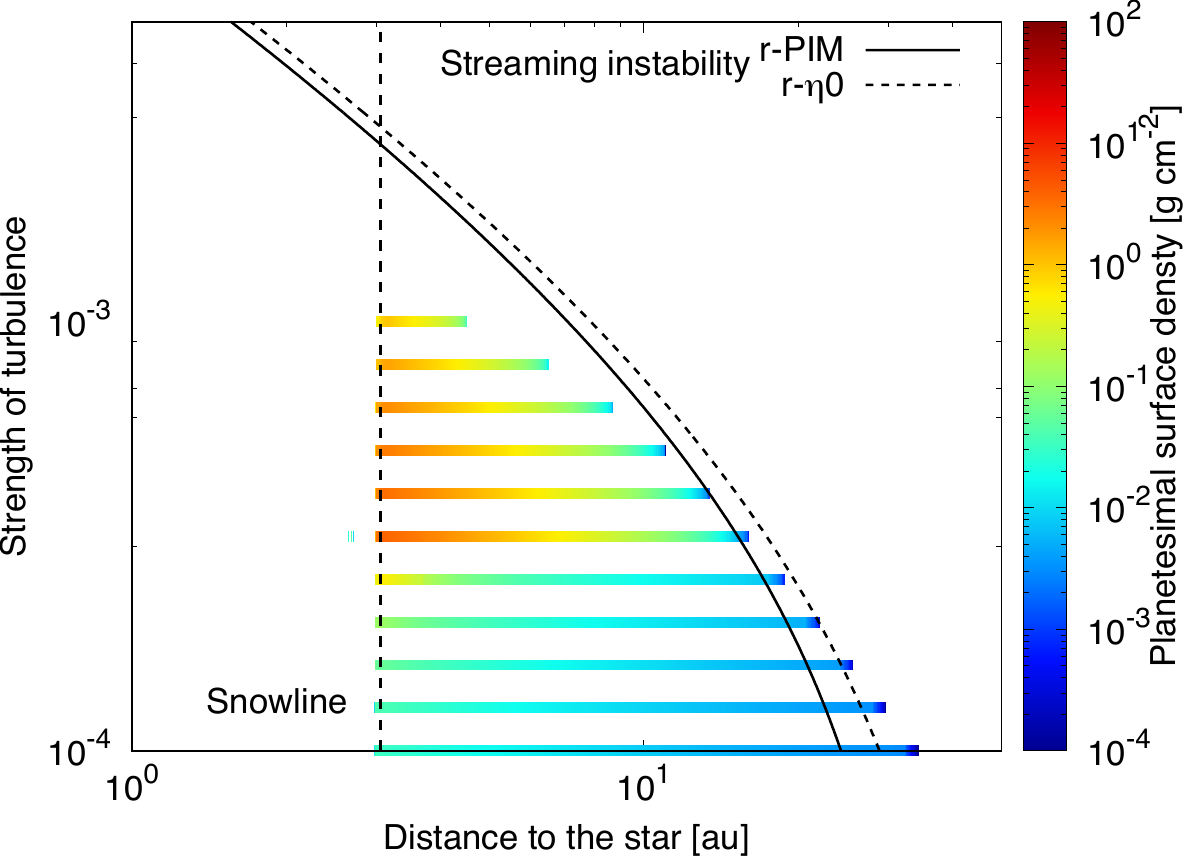}
\includegraphics[width=0.45\linewidth]{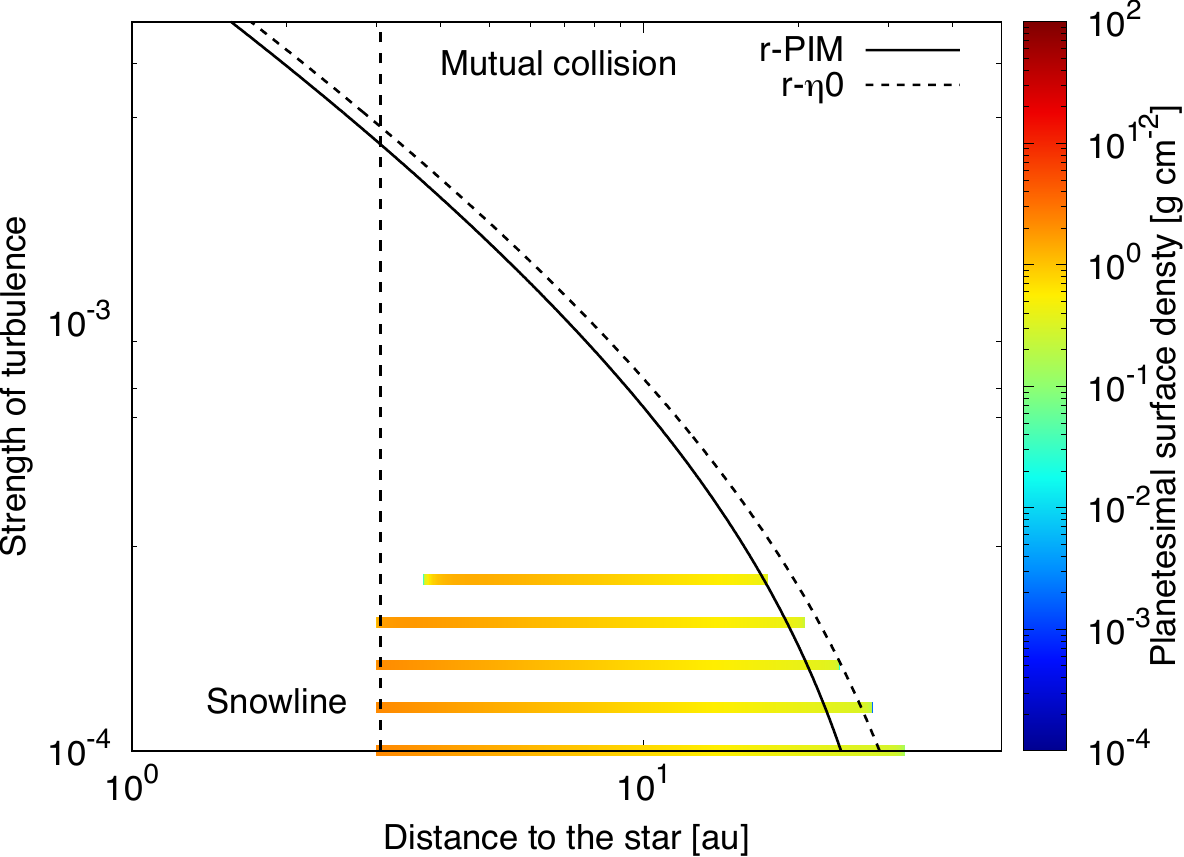}
\caption{Same as Fig. \ref{fig:regions-shorttSI} but with $\tau_{\rm SI}=10^{3}T_{\rm K}$. The left and right panels represent the formation regions of the planetesimals formed by streaming instability and mutual collision, respectively.}
\label{fig:regions-longtSI}
\end{figure*}

\begin{figure}[tbp]
\centering
\includegraphics[width=0.9\linewidth]{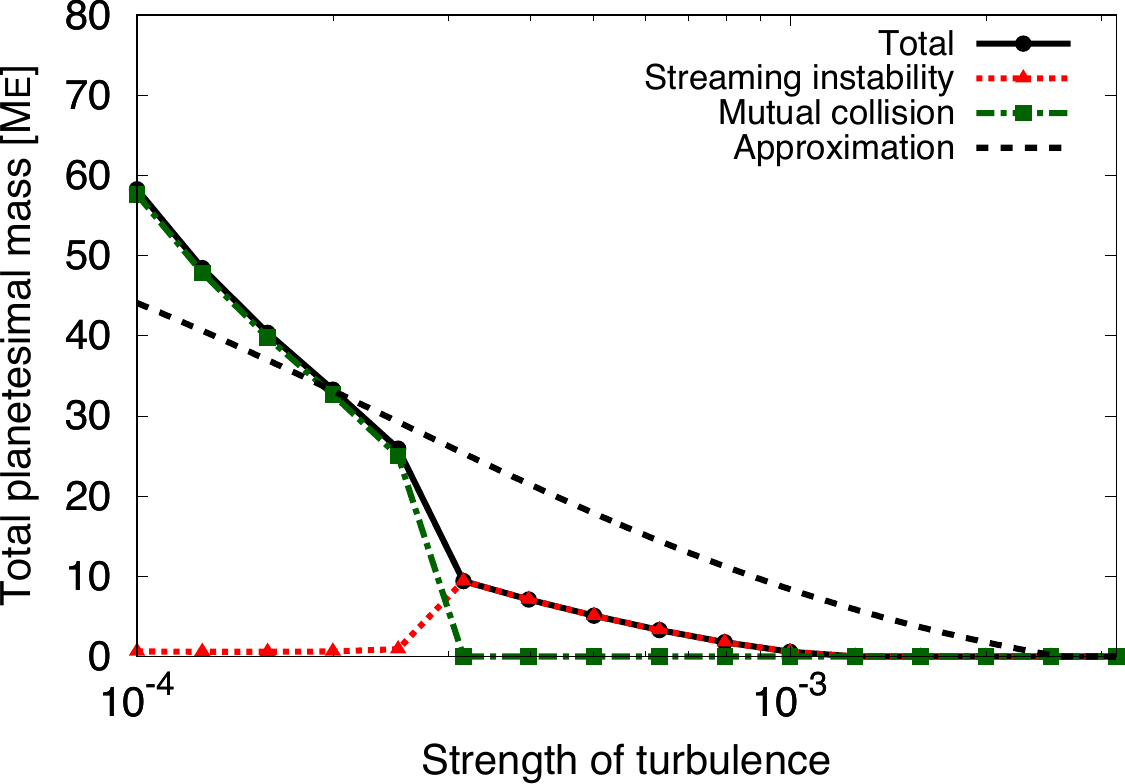}
\caption{Same as Fig. \ref{fig:planetesimal-mass} but with $\tau_{\rm SI}=10^{3}T_{\rm K}$. The red dotted and green dashed-and-dotted curves represent the mass of the planetesimals formed by streaming instability and mutual collision, respectively. The black solid curve represents their total mass.}
\label{fig:planetesimal-mass-longtSI}
\end{figure}

Figure \ref{fig:regions-longtSI} shows that the planetesimal formation region is between $r=r_{\rm SL}$ and $r_{\rm PIM}$, which is the same result with the short SI timescale case including the deviation of the outer edge from $r=r_{\rm PIM}$. The figure also shows that all planetesimals form by streaming instability when $\alpha\geq10^{-3.5}$, but most of the planetesimals form by mutual collision when $\alpha\leq10^{-3.6}$. The left panel also shows that the planetesimal surface density of the outer part of the formation region is smaller than the one with the short SI timescale when $\alpha\geq10^{-3.5}$. This is because only part of the dust drifting into the formation place (i.e., the gas pressure bump) converts to planetesimals, as we explained above. Except for these cases, the surface density of the planetesimals (formed by both mechanisms) are well approximated by Eq. (\ref{Sigmapl_est}) for any strength of turbulence. Figure \ref{fig:planetesimal-mass-longtSI} also shows that the dominant planetesimal formation mechanism is streaming instability when $\alpha\geq10^{-3.5}$ and is mutual collision when $\alpha\leq10^{-3.6}$. When $\alpha\geq10^{-3.5}$, the total mass is much smaller than the approximation by Eq. (\ref{Mplstotest}), because the planetesimal surface density of the outer part of the formation region is smaller than the approximation by Eq. (\ref{Sigmapl_est}).

\subsection{Effects of the Stokes number dependence of streaming instability} \label{St-dependence}
Previous 3D hydrodynamical simulations have shown that the condition and timescale of streaming instability depend on the Stokes number of dust particles. We consider such a case according to the results of \citet{li21}. In this case, the logarithm of the critical density ratio is
\begin{equation}
\log{\epsilon_{\rm crit}}=A(\log{\rm St})^{2}+B\log{\rm St}+C
\label{logepsiloncrit}
\end{equation}with
\begin{equation}
\begin{cases}
A=0, B=0, C=\log{2.5} & {\rm St}\leq0.015, \nonumber \\
A=0.48, B=0.87, C=-0.11 & {\rm St}>0.015.
\end{cases}
\end{equation}
The streaming instability timescale depends on the Stokes number of the particles,
\begin{equation}
\tau_{\rm SI}=
\begin{cases}
\dfrac{2700}{\Omega_{\rm K}} & {\rm St}\leq0.015,\\
\dfrac{40.5}{\rm St~\Omega_{\rm K}} & {\rm St}>0.015,
\end{cases}
\end{equation}
\label{tauSI}
as shown by the approximation of the results of \citet{li21} (see Appendix \ref{clumping}).

Figure \ref{fig:regions-SttSI} represents the surface density and formation regions of planetesimals when streaming instability depends on the Stokes number. The figure shows that the profiles of planetesimals are similar to the case with the short SI timescale (see Fig. \ref{fig:regions-shorttSI}). All planetesimals form by streaming instability, and the planetesimal surface density is well approximated by Eq. (\ref{Sigmapl_est}). The planetesimal formation region lies between $r_{\rm SL}$ and $r_{\rm PIN}$ as well as the other cases. Planetesimals form even when $\alpha=10^{-2.9}$, and the outer edge of the formation region for each $\alpha$ is slightly larger than that with short $\tau_{\rm SI}$. This is because $\epsilon_{\rm crit}$ is smaller than unity when $0.015<{\rm St}(<1)$ (Eq. (\ref{logepsiloncrit})), which is case for the drifting dust (see the third column of Fig. \ref{fig:evolution-dust}).

\begin{figure}
\centering
\includegraphics[width=0.9\linewidth]{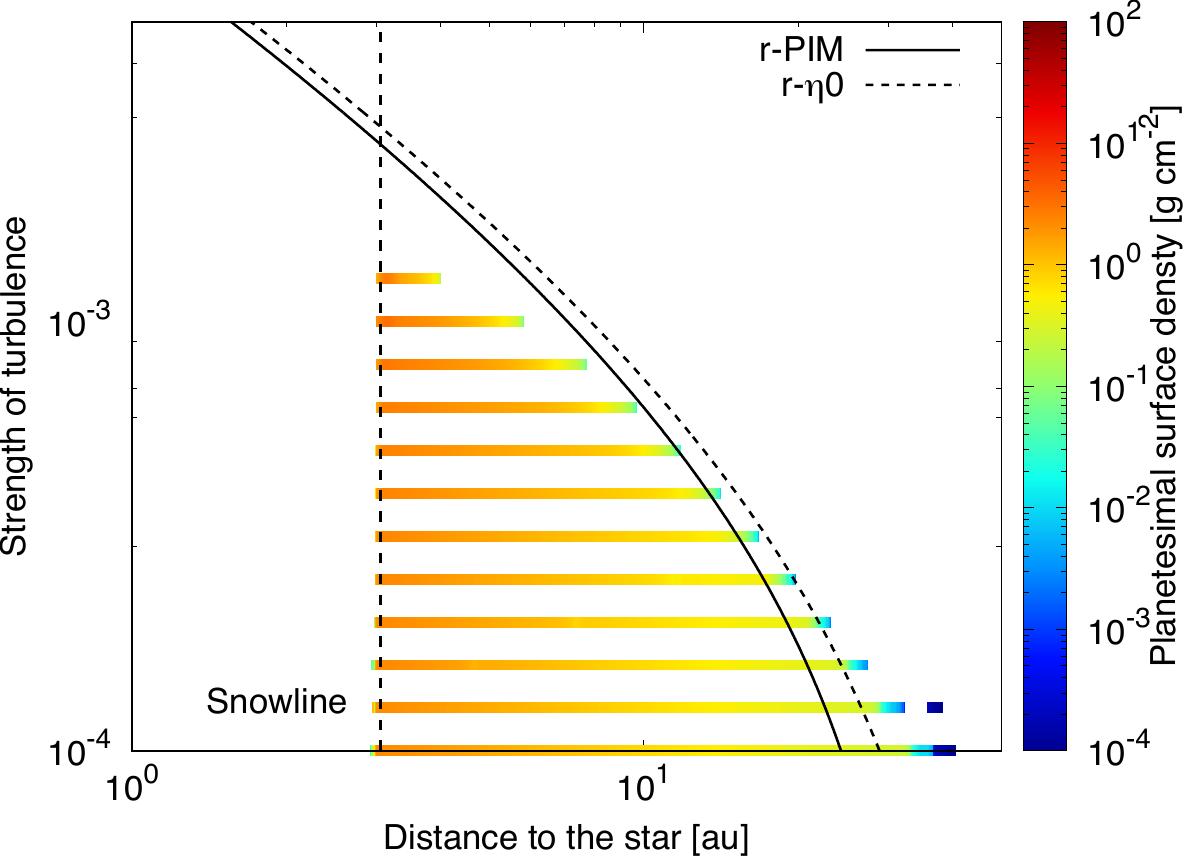}
\caption{Same as Fig. \ref{fig:regions-shorttSI}, but $\epsilon_{\rm crit}$ and $\tau_{\rm SI}$ depend on ${\rm St}$ as Eqs. (\ref{logepsiloncrit}) and (\ref{tauSI}), respectively.}
\label{fig:regions-SttSI}
\end{figure}

\section{Discussion}\label{discussion}
\subsection{Disc properties dependence} \label{dependence}
\begin{figure*}
\centering
\includegraphics[width=0.45\linewidth]{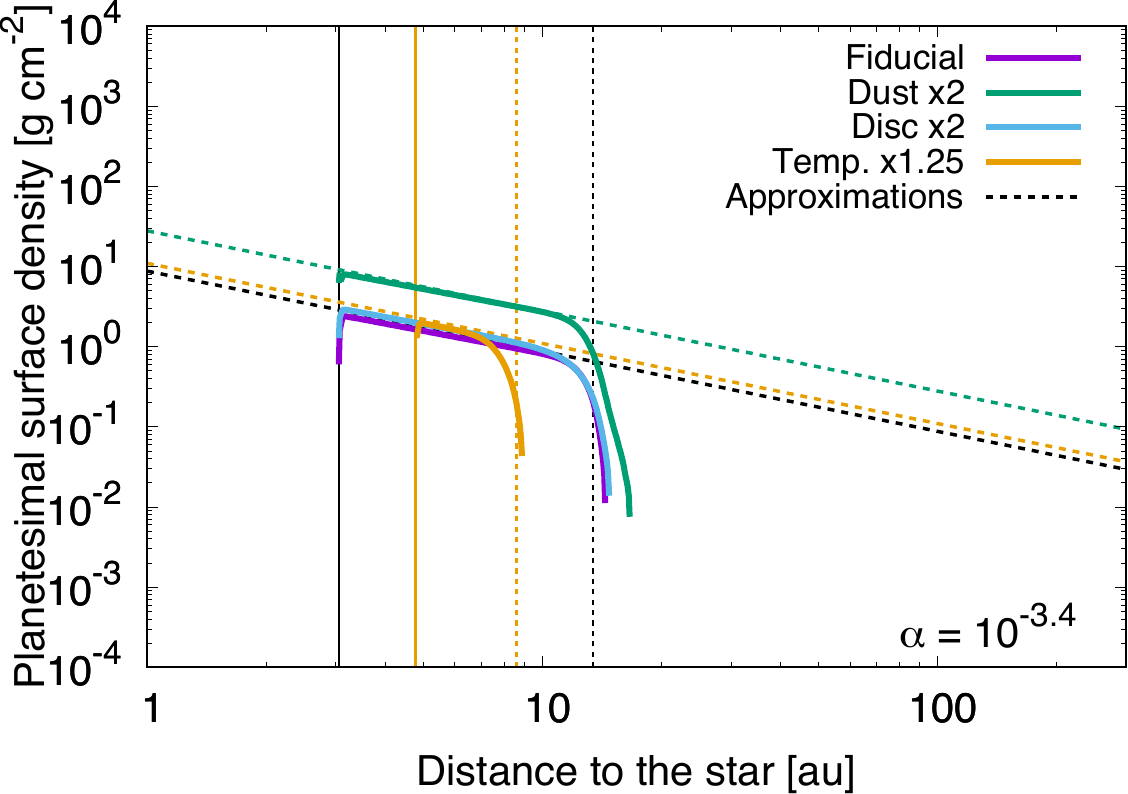}
\includegraphics[width=0.45\linewidth]{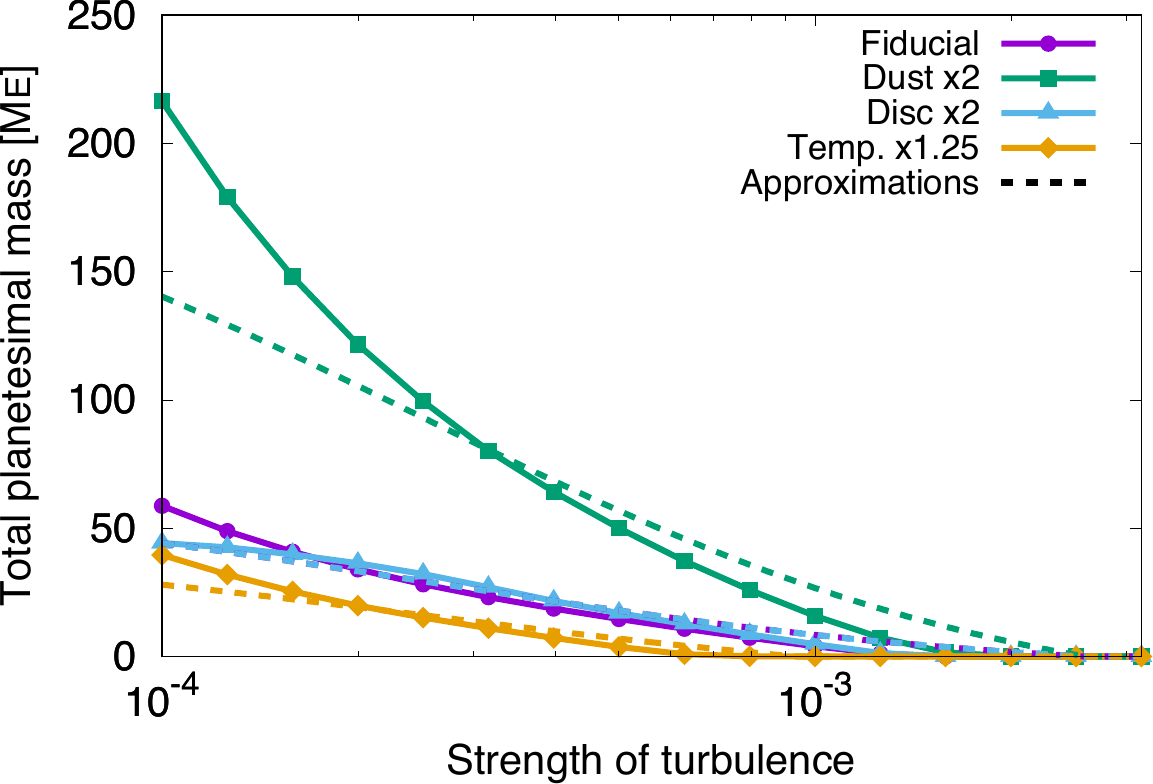}
\caption{Surface density and total mass of planetesimals with various disc properties. The left panel represents the surface density of the planetesimals (solid lines) and corresponding approximations by Eq. (\ref{Sigmapl_est3}) (dotted lines). The vertical solid and dotted lines represent $r_{\rm SL}$ and $r_{\rm PIM}$, respectively. The vertical lines with different $Z_{\Sigma,0}$ and $\Sigma_{\rm g,1au}$ are overlapped with those of the fiducial case. The right panel represents the total planetesimal mass (solid curves) and corresponding approximations by Eqs. (\ref{Mplstotest}) and (\ref{Sigmapl_est3}) (dashed curves).}
\label{fig:planetesimals-various}
\end{figure*}

We investigate the effects of change of disc properties. We consider the cases with various initial dust-to-gas surface density ratio, gas surface density, and disc temperature as described in Table \ref{tab:discs}. In this section, the condition for planetesimal formation by streaming instability is the same with the one used in Section \ref{St-dependence}. Then, we find that planetesimals form perfectly or mainly by streaming instability in any cases. Figure \ref{fig:planetesimals-various} represents the surface density and total mass of the planetesimals (including both planetesimals formed by streaming instability and formed by mutual collision) with various disc properties.

The left panel of Fig. \ref{fig:planetesimals-various} shows the disc properties dependence of the planetesimal surface density. It depends on the dust mass and weakly on the disc temperature but not on the gas disc mass. This dependence can be explained by updating the approximation of planetesimal surface density, Eq. (\ref{Sigmapl_est}). According to \citet{lam14b}, the inward dust mass flux is estimated by the following equation:
\begin{eqnarray}
\dot{M}_{\rm d}=9.5\times10^{-5}\left(\dfrac{\Sigma_{\rm g,1au}}{500~{\rm g~cm}^{-2}}\right)\left(\dfrac{Z_{\Sigma,0}}{0.01}\right)^{5/3} \nonumber \\
\times\left(\dfrac{M_{*}}{M_{\odot}}\right)^{1/3}\left(\dfrac{t}{\rm Myr}\right)^{-1/3}M_{\rm E}~{\rm yr}^{-1},
\label{flux}
\end{eqnarray}
and this is consistent with our result, $\dot{M}_{\rm d}=1.5\times10^{-4}M_{\rm E}~{\rm yr}^{-1}$, when we substitute $t=0.25~{\rm Myr}$ into Eq. (\ref{flux}) (see Appendix \ref{analytical} for more general expressions). Then, by substituting Eq. (\ref{flux}) for Eq (\ref{Sigmapl_est}), we get a general expression:
\begin{eqnarray}
\Sigma_{\rm{pls,est}}
=\dfrac{33.5}{2.728+1.082p}\left(\dfrac{Z_{\Sigma,0}}{0.01}\right)^{5/3}\left(\dfrac{T}{280~{\rm K}}\right)\left(\dfrac{M_{\rm pl}}{20~M_{\rm E}}\right)^{-1} \nonumber \\
\times\left(\dfrac{M_{*}}{M_{\odot}}\right)^{1/2}\left(\dfrac{t}{\rm Myr}\right)^{-1/3}\left(\dfrac{r}{\rm au}\right)^{-1/2}{\rm g~cm}^{-2},
\label{Sigmapl_est2}
\end{eqnarray}
which depends on the dust mass, the disc temperature, and the slope of the gas surface density but not on the disc mass (see also Appendix \ref{analytical} for detailed derivation). In the case of our simulations,
\begin{equation}
\Sigma_{\rm{pls,est}}=8.8\left(\dfrac{Z_{\Sigma,0}}{0.01}\right)^{5/3}\left(\dfrac{T_{\rm 1au}}{280~{\rm K}}\right)\left(\dfrac{r}{\rm au}\right)^{-1}{\rm g~cm}^{-2}.
\label{Sigmapl_est3}
\end{equation}
The left panel of Fig. \ref{fig:planetesimals-various} shows that the obtained planetesimal surface density with various parameters fits the approximation lines (Eq. (\ref{Sigmapl_est3}), dotted lines with corresponding colours) very well. The surface density is $2^{5/3}=3.2$ times higher than that of the fiducial case when $Z_{\Sigma,0}$ is two times higher (green) and is slightly higher when $T_{\rm 1au}$ is 1.25 times higher (orange). On the other hand, the surface density is the same with the one of the fiducial case when $\Sigma_{\rm g,1au}$ is two times higher (light blue).

We also plot the positions of the snowline ($r_{\rm SL}$) and where the planet reaches its pebble-isolation mass ($r_{\rm PIM}$) in the left panel of Fig. \ref{fig:planetesimals-various}. When the disc is hot, $r_{\rm PIM}$ is small ($8.60~{\rm au}$), because the pebble-isolation mass depends on the sound speed (see Eq. (\ref{MPIM})). On the other hand, $r_{\rm PIM}$ is at the same position with the fiducial case ($13.4~{\rm au}$) when the dust or disc mass is changed (black lines). The position of the snowline (where the temperature is $160~{\rm K}$) is changed from the fiducial case ($3.06~{\rm au}$) to $4.78~{\rm au}$ only when the disc temperature is higher.

The right panel of Fig. \ref{fig:planetesimals-various} shows that the total planetesimal mass also depends on the disc properties, and it fits well with the approximation by Eq. (\ref{Mplstotest}) (dotted curves). When $Z_{\Sigma,0}$ is changed, the total mass is in proportion to $\Sigma_{\rm pls,est}$, because the positions of the inner and outer edges of the planetesimal formation region are fixed (see Eq. (\ref{Mplstotest})). Hence, the total mass is $2^{5/3}=3.2$ times heavier when $Z_{\Sigma,0}$ is two times higher (green). As a result, the total planetesimal mass could be about $200~M_{\rm E}$ when $Z_{\Sigma,0}=0.02$ and $\alpha=10^{-4}$. When $\Sigma_{\rm g,1au}$ is large, the total planetesimal mass is the same with the one of the fiducial case, because both surface density and formation region of planetesimals do not depend the gas surface density (light blue). When $T_{\rm 1au}$ is 1.25 times higher, the planetesimal surface density is higher in proportion to the temperature (see Eq. (\ref{Sigmapl_est3})), but the formation region is much narrower (orange). As a result, the total planetesimal mass is smaller than that of the fiducial case.

\begin{table}[tbp]
\caption{Disc properties}
\label{tab:discs}
\centering
\begin{tabular}{llll}
\hline
Cases & $Z_{\Sigma,0}$ & $\Sigma_{\rm g,1au}~[{\rm g~cm}^{-2}]$ & $T_{\rm 1au}~[{\rm K}]$ \\
\hline\hline
Fiducial & 0.01 & 500 & 280 \\
Dust $\times2$ & 0.02 & 500 & 280 \\
Disc $\times2$ & 0.01 & 1000 & 280 \\
Temperature $\times1.25$ & 0.01 & 500 & 350 \\
\hline
\end{tabular}
\end{table}

\subsection{Effects of the planetary growth and the later formation of the planetary core}\label{realistic}
In the previous sections, we have considered the cases with simple assumptions to understand how planetesimals form in belt-like regions. Here, we investigate more realistic situation considering the evolution of the gas disc, later formation of the embedded planet, growth of the planet by gas accretion, and Type II migration of the planet.

In this section, we improve the gas disc model used in the previous sections (Eq. (\ref{gasg})) to express the time and radial reduction of the gas surface density \citep{and09},
\begin{equation}
\Sigma_{\rm g,unp}=\Sigma_{\rm g,1au}\left(\dfrac{r}{\rm au}\right)^{-\gamma}\exp{\left\{-\left(\dfrac{r}{r_{\rm c}}\right)^{2-\gamma}\right\}},
\label{gasg2}
\end{equation}
where $\Sigma_{\rm g,1au}=500\exp{(-t/\tau_{\rm disc})}~{\rm g~cm^{-2}}$ with $\tau_{\rm disc}=3~{\rm Myr}$, $\gamma=1.5-q=1$ (see Eq. (\ref{temperature})), and $r_{\rm c}=150~{\rm au}$. The outer edge of the disc (calculation region) is $300~{\rm au}$ as well as the assumptions in the previous sections.

Planets grown to around the pebble isolation mass also start gas accretion. We consider the growth of the embedded planet by gas accretion as,
\begin{equation}
\dfrac{{\rm d}M_{\rm{pl}}}{{\rm d}t}=\min{\left\{\left(\dfrac{{\rm d}M_{\rm{pl}}}{{\rm d}t}\right)_{\rm KH}, \left(\dfrac{{\rm d}M_{\rm{pl}}}{{\rm d}t}\right)_{\rm disc}, \dot{M}_{\rm g}\right\}},
\label{dMpltot}
\end{equation}
where the first, second, and third terms of the right-hand side represent the gas accretion by the Kelvin–Helmholtz-like contraction of the envelope, the accretion of gas from the protoplanetary disc into the Hill sphere, and the limit due to the global gas accretion rate, respectively \citep{joh19}. The first term is motivated by \citet{iko00},
\begin{equation}
\left(\dfrac{{\rm d}M_{\rm{pl}}}{{\rm d}t}\right)_{\rm KH}=10^{-5}M_{\rm E}~{\rm yr}^{-1}\left(\dfrac{M_{\rm pl}}{10~M_{\rm E}}\right)^{4}\left(\dfrac{\kappa}{1.0~{\rm cm^{2}~g^{-1}}}\right)^{-1},
\label{dMpltot-KH}
\end{equation}
where we assume $\kappa=0.05~{\rm cm^{2}~g^{-1}}$ as the opacity of the envelope. The second term is given by
\begin{equation}
\left(\dfrac{{\rm d}M_{\rm{pl}}}{{\rm d}t}\right)_{\rm disc}=\dfrac{0.29}{3\pi}\left(\dfrac{H_{\rm g,pl}}{r_{\rm pl}}\right)\left(\dfrac{M_{\rm pl}}{M_{*}}\right)^{4/3}\dfrac{\dot{M}_{\rm g}}{\alpha}\dfrac{\Sigma_{\rm g,pl}}{\Sigma_{\rm g,unp}},
\label{dMpltot-disc}
\end{equation}
where $\Sigma_{\rm g,pl}$ is the gas surface density at the planetary orbit inside the gap \citep{tan16}. The global gas accretion rate is \citep{and09}
\begin{equation}
\dot{M}_{\rm g}=3\pi\nu\Sigma_{\rm g,unp}\left\{1-2(2-\gamma)\left(\dfrac{r}{r_{\rm c}}\right)^{2-\gamma}\right\}.
\label{Mdotg}
\end{equation}

As the planet mass increases, the analytical gap model we used in the previous sections is not accurate \citep{duf15a}. Therefore, in this section, we use the model described in Paper 1 (see Appendix \ref{gapmodel} for the details).

The type of the planetary migration also shifts from the Type I to Type II as the gap around the planet is deeper. In order to express the Type II migration as well, we adjust the migration timescale, Eq. (\ref{tmig}), as follows \citep{kan18a}:
\begin{equation}
\tau_{\rm mig,adj}=\tau_{\rm mig}\left(\dfrac{\Sigma_{\rm g,pl}}{\Sigma_{\rm g,unp}}\right)^{-1}=(1+0.04K)\tau_{\rm mig},
\label{tmigadj}
\end{equation}
where the factor $K$ is defined as Eq. (\ref{K}).

The approximation of the planetesimal surface density, Eq. (\ref{Sigmapl_est2}), is then improved,
\begin{eqnarray}
\Sigma_{\rm{pls,est}}
=\dfrac{33.5~(1+0.04K)}{2.728+1.082p}\left(\dfrac{Z_{\Sigma,0}}{0.01}\right)^{5/3}\left(\dfrac{T}{280~{\rm K}}\right)\left(\dfrac{M_{\rm pl}}{20~M_{\rm E}}\right)^{-1} \nonumber \\
\times\left(\dfrac{M_{*}}{M_{\odot}}\right)^{1/2}\left(\dfrac{t}{\rm Myr}\right)^{-1/3}\left(\dfrac{r}{\rm au}\right)^{-1/2}{\rm g~cm}^{-2},
\label{Sigmapl_est_real}
\end{eqnarray}
where we assume the effect of the radial reduction of $\Sigma_{\rm g,unp}$ is negligible.

\begin{figure*}[tbp]
\centering
\includegraphics[width=0.9\linewidth]{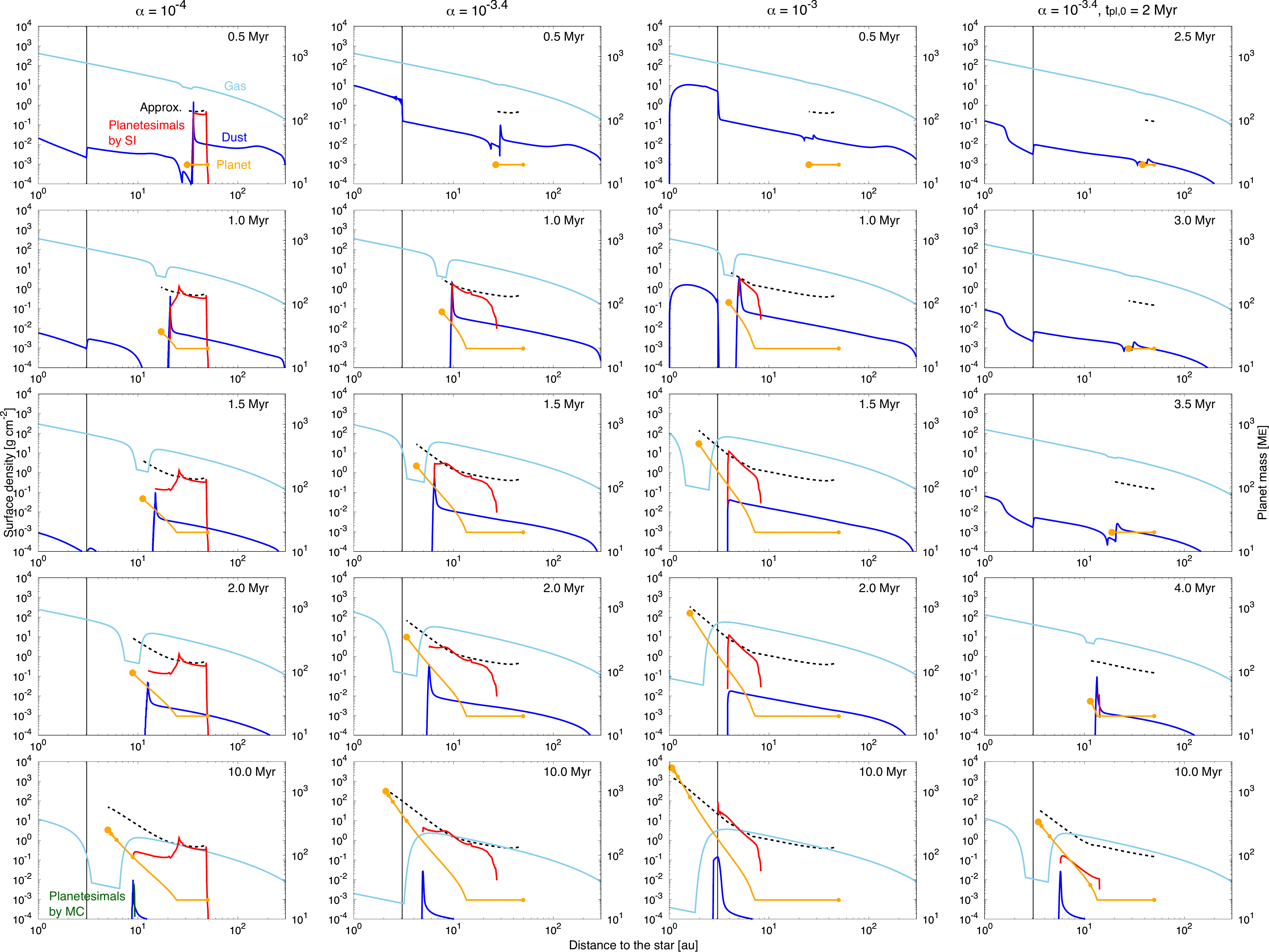}
\caption{Same as Fig. \ref{fig:evolution-Sigma} but considering the evolution of the gas disc, the growth of the embedded planet by gas accretion, and the Type II migration of the planet. The red and green curves represent the surface density of the planetesimals formed by streaming instability and mutual collision, respectively. The mass and orbital position of the planet are also plotted (orange), where the large and small circles represent the mass of that time and of every $2~{\rm Myr}$, respectively. The black dashed curves are the approximation by Eq. (\ref{Sigmapl_est_real}). The first to third columns from the left represent the evolution with $\alpha=10^{-4}$, $10^{-3.4}$, and $10^{-3}$, respectively. The fourth column is the case with $\alpha=10^{-3.4}$, and the planet is put at $t_{\rm pl,0}=2~{\rm Myr}$.}
\label{fig:evolution-Sigma-real}
\end{figure*}

Figure \ref{fig:evolution-Sigma-real} resents the cases where the above effects are included. In all cases, planets grow large and make deep and wide gas gaps during their inward migration. The positions where the planets start to grow are outer as the strength of turbulence is small. After the accretion starts, the migration speed decreases, because the type of migration shifts from the Type I to Type II. Finally, the mass of the planets go to $\sim1000~M_{\rm E}$, and their migration stops. The migration speed before the start of the rapid planetary growth is slower than that of the simple Type I migration case. The migration speed is slower as $\alpha$ is smaller, because the gap is deeper (Eq.(\ref{tmigadj})). After the rapid growth starts, the migration speed does not depend on $\alpha$ so much, because the gas accretion rate is small as the turbulence is strong (Eq. (\ref{dMpltot-disc})), which cancels out the above effect.

The first column shows that planetesimals form (by streaming instability) from the start of the calculation, where the planetary orbit is $50~{\rm au}$ (see the top panel at $t=0.5~{\rm Myr}$). This outer edge of the formation region is farther than that in the previous sections, because the gap model used in this section is different from the other sections' one. The surface density of the formed planetesimals is well reproduced by the updated approximation by Eq (\ref{Sigmapl_est_real}) (the black dashed curve) and continuously increases as the planet migrates inward. Then, the pebble front reaches the outer edge of the disc by $1.0~{\rm Myr}$, and the inward dust flux decreases, resulting in the rapid reduction of the planetesimal surface density at $26~{\rm au}$ (the second top panel), which is not expressed in the approximation. After that, planetesimal formation continues until the stop of the migration of the planet ($2.0-10.0~{\rm Myr}$). Small amount of planetesimals also form by mutual collision at this stage (green). Since the planet stops before it reaches the snowline, the inner edge of the planetesimal formation region is at $9~{\rm au}$, which is outer than that of our previous results without the planetary growth and the Type II migration.

The second column shows the case where $\alpha=10^{-3.4}$. Planetesimals start to form at $27~{\rm au}$, which is inner than that with $\alpha=10^{-4}$ (the top panel). This trend of the position of the outer edge of the planetesimal formation region is the same with the cases without the planetary growth and the Type II migration. At $13~{\rm au}$, the slope of the surface density of planetesimals starts to be steeper than that without the additional effects ($\Sigma_{\rm pls}\propto r^{-1}$), because the planet starts the rapid growth, and so the migration speed becomes slow (the second top). This change of the slope is well reproduced by the approximation (Eq. (\ref{Sigmapl_est_real})) showing that the planetesimal surface density is proportional to $M_{\rm pl}r^{-1}$ when $K\gg25$ and $T\propto r^{-1}$. However, the increase of the planetesimal surface density stops at $8~{\rm au}$, because the dust inward flux decrease after the pebble front reaches the outer edge of the disc. Finally, the planet crosses the snowline, but the outer edge of the gap (i.e., the gas pressure maximum) is still outside the snowline, and the inner edge of the planetesimal formation region is outside the snowline as well although it is inner than the case with $\alpha=10^{-4}$.

The third column shows the profiles with $\alpha=10^{-3}$. The start of planetesimal formation is later than that with weaker turbulence, when the pebble front has already reached the outer edge of the disc and the inward dust flux has decreased (the second top panel). Since the planet rapidly grows by gas accretion and the migration speed decreases, the slope of the planetesimal surface density is steeper than that without the planetary growth and the Type II migration, which is well reproduced by the approximation (Eq. (\ref{Sigmapl_est_real})). The planet migrates inward to $1~{\rm au}$, where the pressure maximum also reaches the snowline (the bottom panel). The inner edge of the planetesimal formation region is then at the snowline as well as the cases without the planetary growth and the type II migration.

The fourth column shows the case where $\alpha=10^{-3.4}$ and the planetary core forms at $t_{\rm pl,0}=2.0~{\rm Myr}$. The top panel shows the properties when the time has passed already $2.5~{\rm Myr}$. Since the amount of the gas and dust still exist in the disc is smaller than the other cases, the planetary growth is less efficient and the dust inward flux is smaller. As a result, the start position of the planetesimal formation (i.e., the outer edge of the planetesimal formation region) is inner than the case with the earlier formation of the planetary core (the second column), and the planetesimal surface density is smaller. The planetesimal surface density is also smaller than the approximation which does not consider the decrease of the inward dust flux due to the pebble front reaches the outer edge of the disc. Also, the planet migrates to only just outside the snowline, resulting in the inner edge of the planetesimal formation region is outer than the the earlier planetary core formation case.

\begin{figure}[tbp]
\centering
\includegraphics[width=0.9\linewidth]{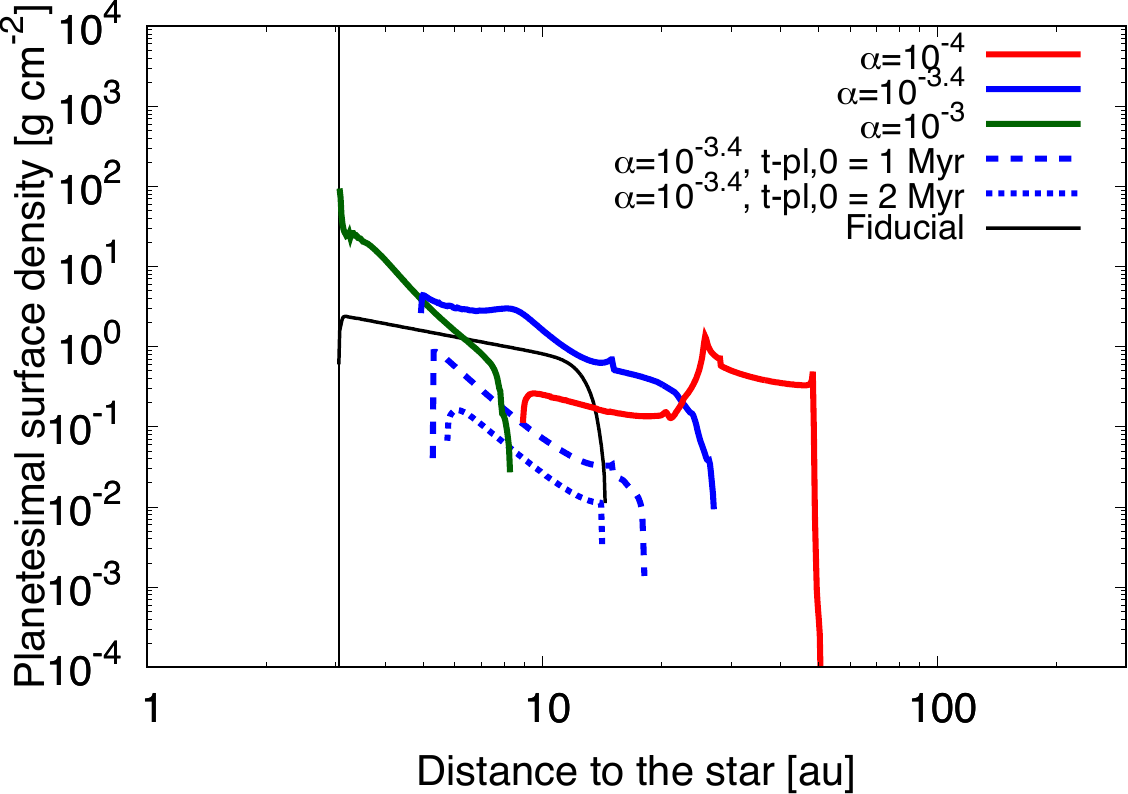}
\caption{Final profiles of the total planetesimal surface density considering the evolution of the gas disc, the growth of the embedded planet by gas accretion, and the Type II migration of the planet. The red, blue, and green solid blue curves represent the profiles with $\alpha=10^{-4}$, $10^{-3.4}$, and $10^{-3}$, respectively. The solid, dashed and dotted curves are the cases, where the planet is put at $t_{\rm pl,0}=0~{\rm Myr}$, $1~{\rm Myr}$, and $2~{\rm Myr}$, respectively, with $\alpha=10^{-3.4}$. The black curve is the case without the additional effects when $\alpha=10^{-3.4}$ (the same as the Fiducilal case in the left panel of Fig. \ref{fig:planetesimals-various}).}
\label{fig:planetesimals-real}
\end{figure}

Figure \ref{fig:planetesimals-real} shows the final distribution of the planetesimal surface density. As we explained in the previous paragraphs, the growth of the planet makes its migration slower, which changes the profiles from those without the growth (the black curve). Due to the slower migration, the embedded planet stops at the point where the gas pressure bump has not reached the snowline yet, which makes the inner edge of the planetesimal formation region outer. The timing of the pebble front reaches the outer edge of the disc (i.e., the inward dust mass flux decreases) is also an important factor for the profiles of the belt-like planetesimal formation region. Also, if the formation of the embedded planet is late, the dust inward flux has already been small, resulting in the low planetesimal surface density and the narrow planetesimal formation region.

\begin{figure*}[tbp]
\centering
\includegraphics[width=0.45\linewidth]{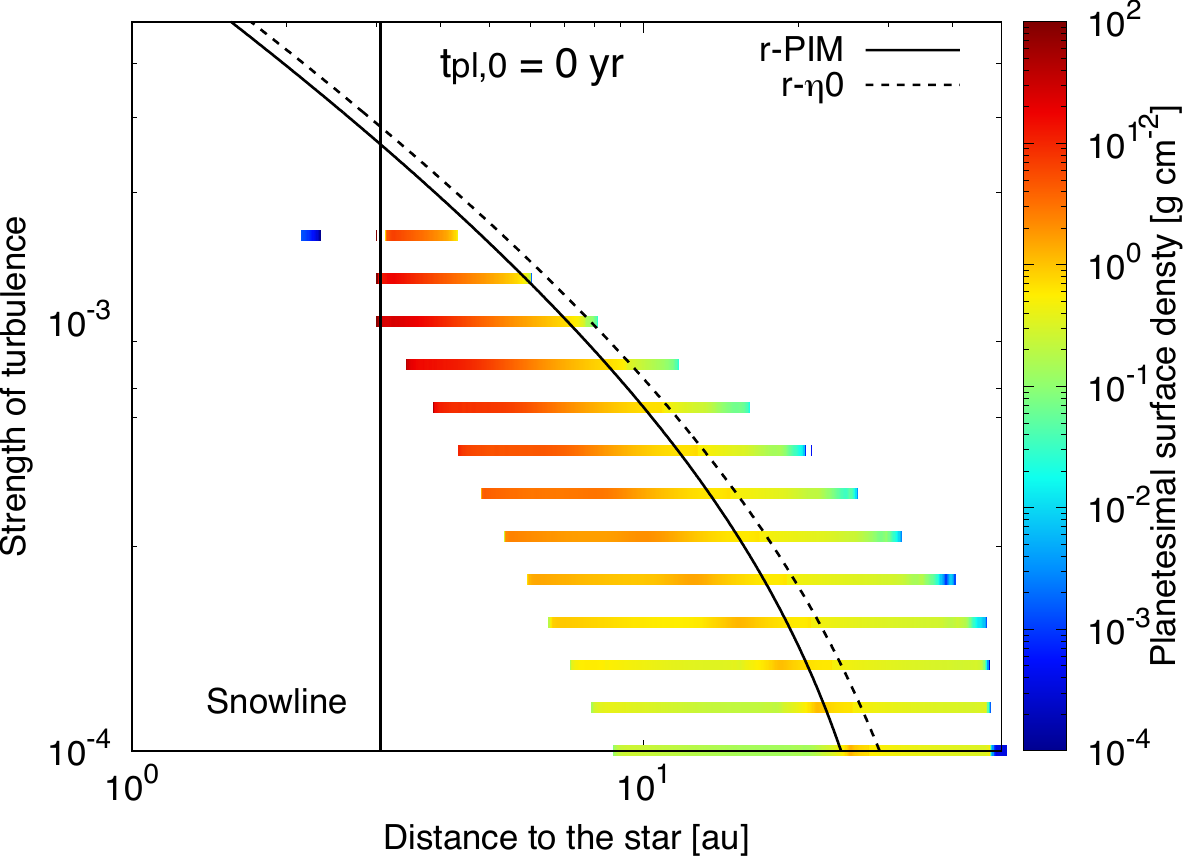}
\includegraphics[width=0.45\linewidth]{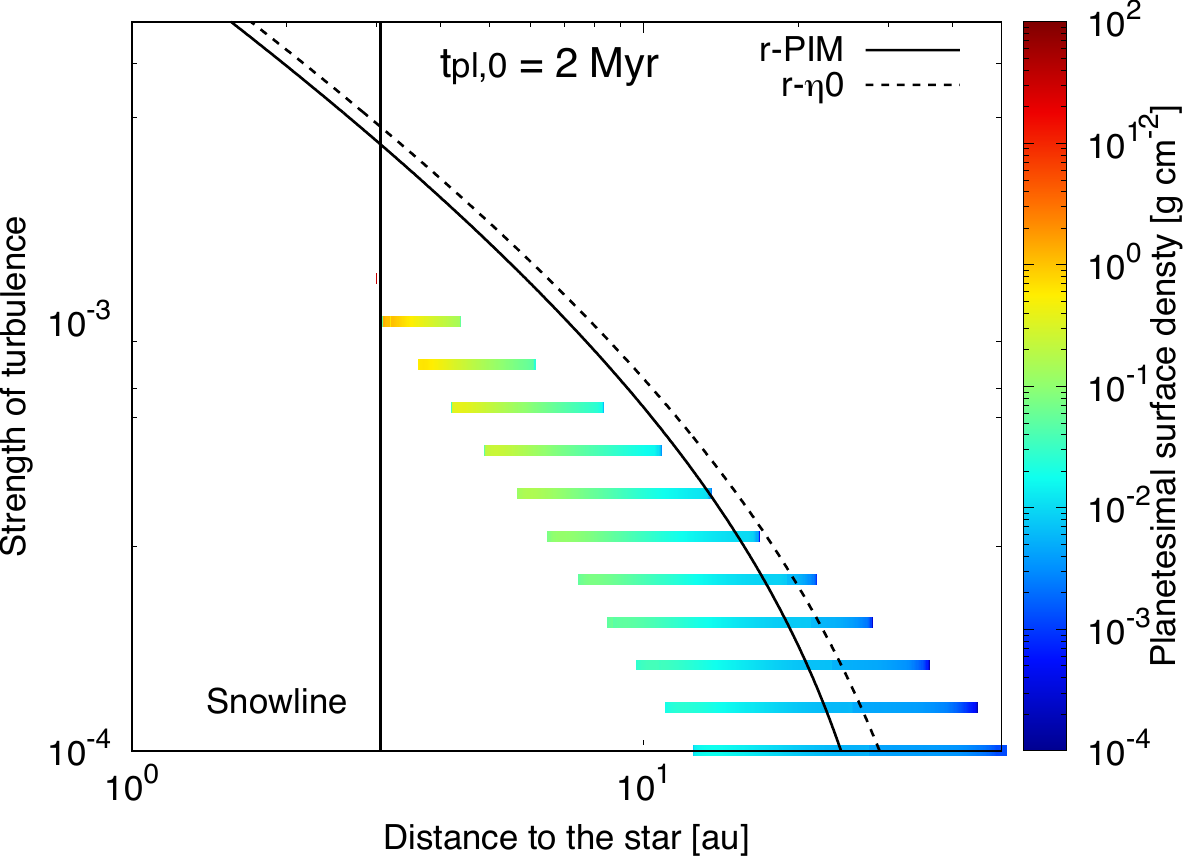}
\caption{Same as Fig. \ref{fig:regions-shorttSI} but considering the evolution of the gas disc, the growth of the embedded planet by gas accretion, and the Type II migration of the planet. The left and right panels represent the cases where $t_{\rm pl,0}=0$ and $2~{\rm Myr}$, respectively. The black curves are the same with those in Fig. \ref{fig:regions-shorttSI} assuming that the planet mass and gas disc are fixed.}
\label{fig:regions-real}
\end{figure*}

Figure \ref{fig:regions-real} also shows that the belt-like planetesimal formation region is formed with the planetary growth. In the case where the planet exists from the start of the calculation (the left panel), the planetesimal formation region spreads outside $r_{\rm PIM}$, the orbital position where the planet reaches the pebble isolation mass (same with the previous sections' one; the planet mass, the gas disc, and the Stokes number of the dust are fixed), farther than the fiducial case in the previous sections. This is simply because the gas gap model we use here is different from the previous one. The inner edge of the planetesimal formation region is outside the snowline, because the migration speed of the planet decreases as the planet grows heavy, and the gas disc disappears before the outer edge of the gap reaches the snowline. The $\alpha$ dependence of the orbital position of the inner edge of the formed planetesimal belt reflects the $\alpha$ dependence of the migration speed as we discussed in the previous paragraphs. The panel also shows that the slope of the planetesimal surface density is gentle at the inner part of the belt, or even a peak is formed when $\alpha\leq10^{-3.4}$, which is because the pebble front reaches the outer edge of the disc and then the dust mass flux flowing into the planetesimal formation place decreases.

The right panel of Fig. \ref{fig:regions-real} shows that the planetesimal formation region with $t_{\rm pl,0}=2~{\rm Myr}$ has similar $\alpha$ dependence to that with $t_{\rm pl,0}=0~{\rm Myr}$ (the left panel), but it is narrower, and the value of the planetesimal surface density is much lower. This is because the pebble front has already reached the outer edge of the disc, and the inward dust mass flux has decreased, as we interpreted Figs. \ref{fig:evolution-Sigma-real} and \ref{fig:planetesimals-real} in the previous sections.

\begin{figure}[tbp]
\centering
\includegraphics[width=0.9\linewidth]{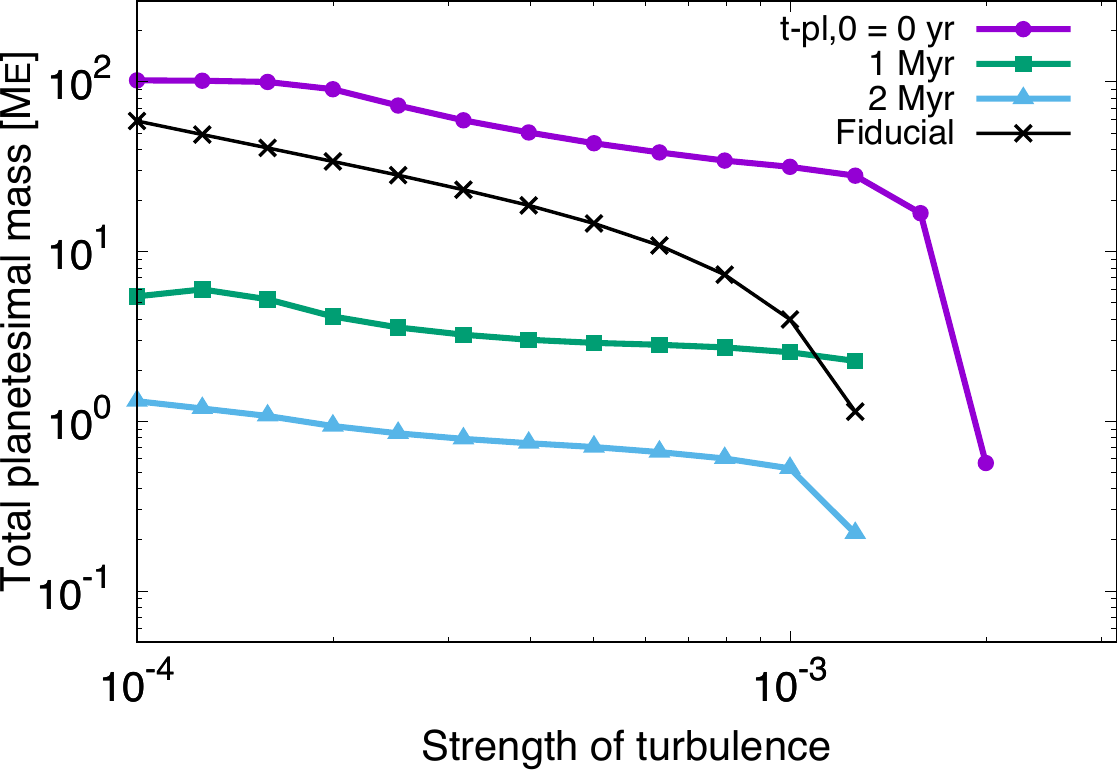}
\caption{Same as Fig. \ref{fig:planetesimal-mass} but considering the evolution of the gas disc, the growth of the embedded planet by gas accretion, and the Type II migration of the planet. The vertical axis is in the logarithmic scale. The purple, green, and sky blue curves represent the total planetesimals mass where the planet is put at $t_{\rm pl,0}=0~{\rm yr}$, $1~{\rm Myr}$, and $2~{\rm Myr}$, respectively. The black curve is the case without the additional effects when $\alpha=10^{-3.4}$ (the same as the Fiducilal case in the right panel of Fig. \ref{fig:planetesimals-various}).}
\label{fig:planetesimal-mass-real}
\end{figure}

Figure \ref{fig:planetesimal-mass-real} shows that the total mass of the formed planetesimals can be $30-100~M_{\rm E}$ when the planet is formed at $t_{\rm pl,0}=0~{\rm yr}$ (the purple curve). This is higher than the cases without the planetary growth and the Type II migration (the black curve), because the planetesimal formation starts at outer orbital positions than those in the previous section due to the change of the gas gap model. On the other hand, the total mass where the planet is formed at $t_{\rm pl,0}=1~{\rm Myr}$ (green) and $2~{\rm Myr}$ (sky blue) is about 10 and 100 times smaller then that with $t_{\rm pl,0}=0~{\rm yr}$, respectively. This is because the planetesimal surface density is smaller and the widths of the planetesimal formation regions are narrower as we discussed in the above paragraph (Fig. \ref{fig:planetesimals-real}). We note that Fig. \ref{fig:regions-real} shows that the planetesimal formation region may spread farther than $r=50~{\rm au}$, and the planetesimal mass may be heavier than $100~M_{\rm E}$, when $\alpha\leq10^{-3.7}$ and $t_{\rm pl,0}=0~{\rm yr}$ if the planet forms farther than our assumption.

\subsection{Effects of the back-reaction from dust to gas}\label{back-reaction}
We do not consider the effects of the back-reaction from dust to gas, which could change the gas structure at the pressure bump and prevent the accumulation of dust. A gas and dust 2D (radial and vertical) hydrodynamical simulation by \citet{tak16} shows that the deformation of the gas pressure bump by the back-reaction prevents direct gravitational instability, and the size of the planetesimals formed by streaming instability becomes smaller even if they form. However, a 2D simulation including the stellar vertical gravity shows that gravitational instability occurs at the gas pressure bump \citep{oni17}. A gas and dust 2D (radial and azimuth) hydrodynamical simulation with a fixed-orbit planet including a simple dust growth model by \citet{kan18b} shows that the back-reaction makes the gas pressure bump flatter, and extreme dust accumulation is suppressed. However, the dust-to-gas density ratio in the midplane $Z_{\rho}$ is still about unity, which satisfies the condition for planetesimal formation by streaming instability. A similar simulation but with a migrating planet and a fixed size of dust by \citet{kan21b} shows that the gas pressure bump does not constantly follow the inward migration of the planet, and multiple dust rings form when $\alpha\leq3\times10^{-4}$. If this occurs even when the dust growth and the planetesimal formation are considered, the radial profile of the planetesimal surface density inside the formation region will be different from our results. On the other hand, a recent gas and dust 3D hydrodynamical simulation with a fixed size of dust by \citet{bi23} shows that, although the accumulation is moderate when $Z_{\rho}>1$, the dust ring is narrower rather than wider when $Z_{\rho}<1$, which is different from the results of the previous 2D simulations. Therefore, more precise 3D simulations considering the dust growth with a migrating planet should be conducted in order to predict the precise formation process of the planetesimals in the future.

\subsection{Effects of the dust leak}\label{dustleak}
We use a dust evolution model expressing the dust mass at each distant as single peak (largest) mass. This is consistent with full-size simulations because the mass is dominated by the peak (largest) mass. However, the dust has mass distribution in reality, and small dust is relatively easy to escape from the accumulation at the gas pressure bump due to the diffusion. This is obvious from the condition for the accumulation of dust, which is determined by the ratio of the speeds of diffusion and drift \citep{zhu12} (see Eqs. (\ref{vdrift}) and (\ref{vdiff})),
\begin{equation}
\left|\dfrac{v_{\rm drift}}{v_{\rm diff}}\right|\sim\dfrac{\rm St}{\alpha}\left(\dfrac{\partial\ln{\Sigma_{\rm g}}}{\partial\ln{r}}\right)\left(\dfrac{\partial\ln{Z_{\rm\Sigma}}}{\partial\ln{r}}\right)^{-1}.
\label{dSigmatot}
\end{equation}
Hence, if the fragmentation of the piled-up dust is efficient, and a lot of small dust is formed, the gas pressure bump may not be able to maintain the accumulation of dust.

However, a recent full-size simulation by \citet{sta23} shows that small particles leak from the gas pressure bump, but the dust-to-gas surface density ratio maintains $Z_{\Sigma}\gtrsim0.01$ for $\sim1~{\rm Myr}$ when the critical fragmentation speed is $v_{\rm cr}=10~{\rm m~s^{-1}}$, which is sufficient for triggering streaming instability outside the snowline \citep{li21}. Therefore, our scenario of planetesimal formation should still work even if the leak of small dust is considered. A recent 2D gas and fixed-sized dust shearing-box simulation by \citet{lee22} shows that dust particles smaller than ${\rm St}\sim0.1$ cannot pile up at the gas pressure bump, and the trap efficiency is $\sim80\%$ even for the particles with ${\rm St}\gtrsim0.1$. This result suggests that the surface density and total mass of actually formed planetesimals could be smaller than our results, but a significant leak of dust should not happen due to the quick growth of dust at the gas pressure bump predicted in our simulations.

\subsection{Effects of the vertical stirring by the planet} \label{vertical}
Recent 3D simulations show that an embedded planet vertically stirs dust settling onto the midplane \citep[e.g.][]{bin21}. This effect can reach the outer region of the protoplanetary disc as the planet is heavy. Here, we briefly check how this vertical stirring changes our results by mimicking the situation.

\begin{figure}
\centering
\includegraphics[width=0.9\linewidth]{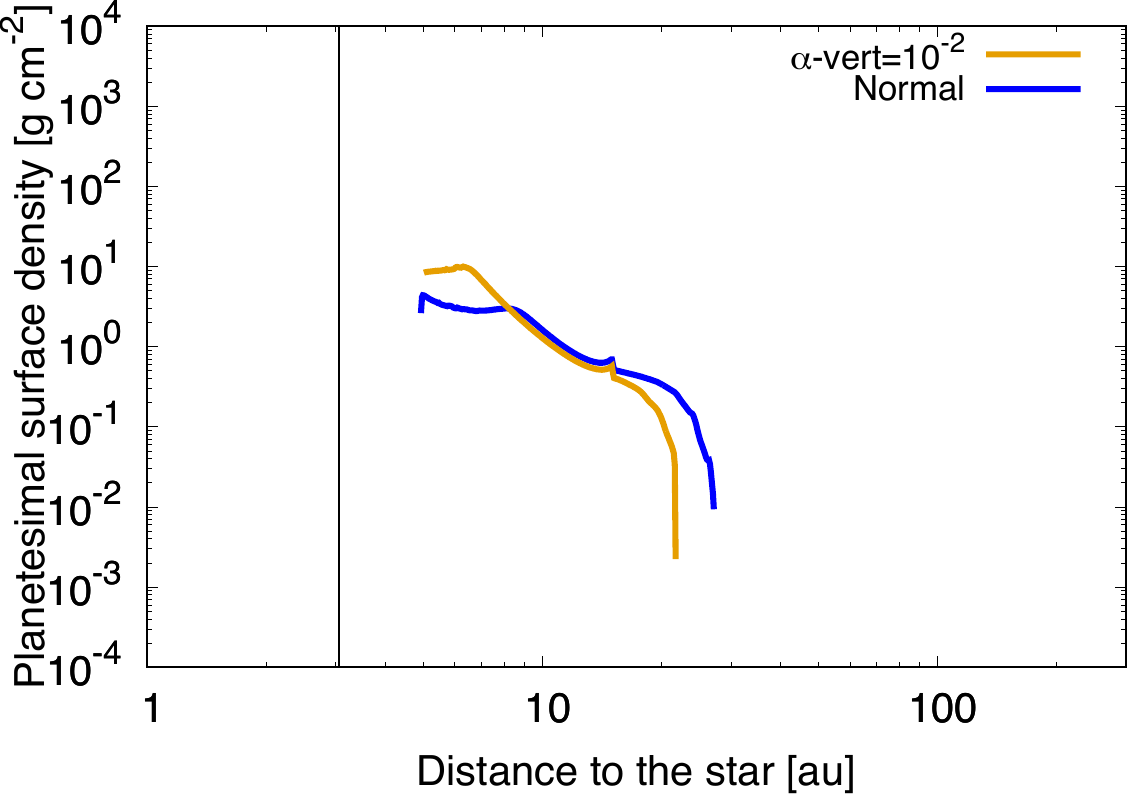}
\caption{Same as Fig. \ref{fig:planetesimals-real}, but the orange curve is a vertically stirred situation, $\alpha_{\rm vert}=10^{-2}$ and $\alpha=10^{-3.4}$. The blue curve is the normal case with $\alpha=10^{-3.4}$, the same with the blue curve in Fig. \ref{fig:planetesimals-real}.}
\label{fig:planetesimals-vertical}
\end{figure}

We treat the strength of turbulence $\alpha$ in Eq. (\ref{Hd}), dominating the vertical equilibrium distribution, as $\alpha_{\rm vert}=10^{-2}$. We assume this value is spatially and temporally constant. Figure \ref{fig:planetesimals-vertical} shows that the final distribution of the planetesimal surface density is similar to the normal case. The total formed planetesimal mass is $59~M_{\rm E}$, which is also similar to that of the normal case, $52~M_{\rm E}$. In the case of the vertical stirring, however, the starting point of the planetesimals formation (i.e., the outer edge of the planetesimal formation region) is inner than the normal case. This is because, the dust density in the midplane is lower than that of the normal case because of the vertical stirring, which makes the timing that the condition for the streaming instability is satisfied later. This trend is also shown in Fig. \ref{fig:rhodgmax}, where $\alpha$ is changed. At the inner part of the planetesimal distribution, the surface density of the vertical stirring case is higher than that of the normal case. This is because the timing of the pebble front reaches the disc outer edge is later than the normal case due to the less efficient collisional growth of dust on the midplane. The vertical stirring makes the midplane dust density lower, and the collision rate becomes smaller. However, the vertical stirring is weaker as the distance to the planet is farther, the speed of the pebble front may not change so much, especially when the planet (or planetary core) is not heavy. On the other hand, the difference of the starting points of the planetesimal formation will remain in the case of a lighter planet.

The presence of an embedded planet may also make the dust ring wider. It will change the starting position of the planetesimal formation and may make the planetesimal surface density lower if the planetesimal formation rate is also affected by the planet. However, the big picture of the planetesimal formation mechanism will not change, similar to the effects of the dust buck reaction (see Section \ref{back-reaction}).

\section{Conclusions} \label{conclusions}
A planet carves the protoplanetary disc and creates a gas pressure bump. Dust drifting from the outer region of the disc piles up there and form planetesimals. As the planet migrates inward, the planetesimal formation place also moves inward, and the formation region spreads on the inner disc. As a result, planetesimals form in a wide belt-like region in the disc \citep{shi20}.

We investigated this scenario for planetesimal formation by considering in addition the global dust evolution in a protoplanetary disc with a wide range of the value of $\alpha$, the strength of turbulence. We showed that the dust particles pile up at the bump and form planetesimals by streaming instability and mutual collision. As the planet migrates inward, the formation region lies roughly between the snowline and the orbital position where the planet reaches its pebble-isolation mass when $10^{-4}\leq\alpha\leq10^{-3}$, which is broadly consistent with observed value of $\alpha$ \citep{pin22}. As $\alpha$ is smaller, the planetesimal formation region is wider, and the total mass of the formed planetesimals is heavier.

The formation mechanism depends on the SI (streaming instability) timescale. In the case of short SI timescale, all planetesimals form by streaming instability independent from the value of $\alpha$. On the other hand, in the case of long SI timescale, all planetesimals form by streaming instability when $\alpha\geq10^{-3.5}$, but most of them form by mutual collision when $\alpha\leq10^{-3.6}$. We also investigated the case that the condition for streaming instability and the SI timescale depend on the Stokes number of dust \citep{li21}. The results are almost the same with those with the short SI timescale except for that the outer edge of the planetesimal formation region is slightly farther out.

The planetesimal surface density is $\sim10~{\rm g~cm^{-2}}$ around the inner edge and $\sim0.1-1~{\rm g~cm^{-2}}$ around the outer edge of the formation region. This is consistent with the results of \citet{shi20}, and that means almost all dust drifting into the formation place is converted immediately to planetesimals at the place. The total planetesimal mass depends on $\alpha$, and it reaches about $60~M_{\rm E}$ with $\alpha=10^{-4}$ and typical initial dust-to-gas surface density ratio (i.e., dust mass). Also, the total planetesimal mass depends on the dust mass significantly, and it can be about $200~M_{\rm E}$ when the initial dust-to-gas surface density ratio is $0.02$. We also showed that the surface density and total mass of the planetesimals can be approximated with simple expressions.

Furthermore, when the growth of the embedded planet by gas accretion and the shift to the Type II migration are considered, the profiles of the planetesimal surface density change from those with the simple assumptions. The slowdown of the migration of the planet makes the slopes of the profiles steeper at their inner regions, but it also reduces the surface density once the pebble front reaches the outer edge of the disc and the dust (pebble) inward mass flux decreases. When $10^{-4}\leq\alpha\leq10^{-3}$, the total mass of the formed planetesimals is about $30-100~M_{\rm E}$ if the planetary core has already existed at $t=0~{\rm yr}$. However, the total mass is about 10 and 100 times smaller in the cases where the planetary core forms at $t=1~{\rm Myr}$ and $2~{\rm Myr}$, respectively, because the most of the dust (pebbles) has already fallen to the star before the planetesimal formation starts.

\begin{acknowledgements}
We thank the referee, Joanna Dr{\k{a}}{\.z}kowska, for the very valuable comments. We also thank Christoph Mordasini and Takahiro Ueda for constrictive and useful discussion. This work has been carried out within the framework of the NCCR PlanetS supported by the Swiss National Science Foundation under grants 51NF40\_182901 and 51NF40\_205606.
\end{acknowledgements}

\bibliographystyle{aa}
\bibliography{Shibaike_pls2}

\begin{appendix}
\section{Clumping timescale of dust} \label{clumping}
\begin{figure}[tbp]
\centering
\includegraphics[width=0.9\linewidth]{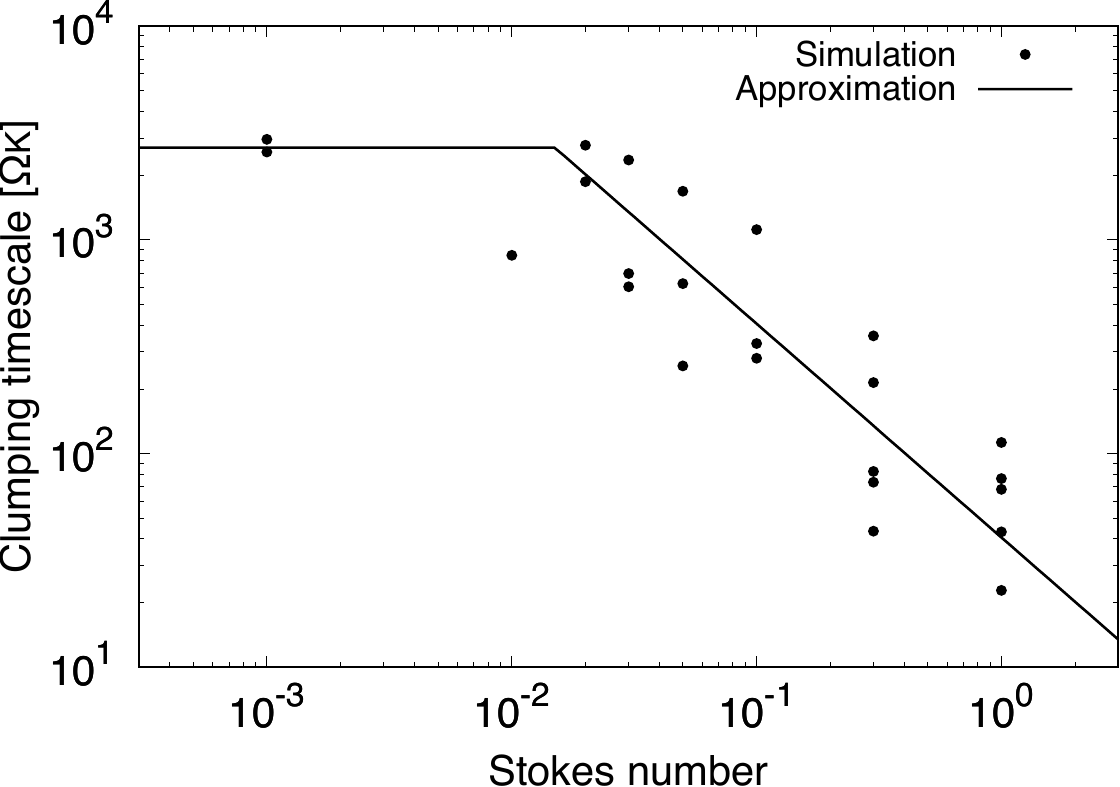}
\caption{Clumping timescale of dust with various value of Stokes number. The circles represent the simulation results of \citet{li21} ($\tau_{\rm S}$ in Table 1 and $t_{\rm pre-cl}$ in Table 2 of the paper). The solid curve represents our approximation of the results (Eq. (\ref{tauSI})).}
\label{fig:clumping-timescale}
\end{figure}

Clumping timescale of dust, in other words, the SI timescale depends on the Stokes number of dust particles. We approximate (by eye) the results of a recent vertically stratified gas and dust hydrodynamical simulation by \citet{li21}. Figure \ref{fig:clumping-timescale} shows the results of the work and our approximation of the results, Eq. (\ref{tauSI}).

\section{Analytical explanation of the approximate expressions} \label{analytical}
The analytical expression of the inward dust (pebble) flux provided by \citet{lam14b} can be expressed as a more general expression:
\begin{eqnarray}
\dot{M}_{\rm d}=9.5\times10^{-5}\left(\dfrac{\Sigma_{\rm ump,g}}{500~{\rm g~cm}^{-2}}\right)\left(\dfrac{Z_{\Sigma,0}}{0.01}\right)^{5/3} \nonumber \\
\times\left(\dfrac{M_{*}}{M_{\odot}}\right)^{1/3}\left(\dfrac{t}{\rm Myr}\right)^{-1/3}\left(\dfrac{r}{\rm au}\right)M_{\rm E}~{\rm yr}^{-1}.
\label{flux-A}
\end{eqnarray}
When $\Sigma_{\rm ump,g}=\Sigma_{\rm g,1au}(r/{\rm au})^{-p}$,
\begin{eqnarray}
\dot{M}_{\rm d}=9.5\times10^{-5}\left(\dfrac{\Sigma_{\rm 1au}}{500~{\rm g~cm}^{-2}}\right)\left(\dfrac{Z_{\Sigma,0}}{0.01}\right)^{5/3} \nonumber \\
\times\left(\dfrac{M_{*}}{M_{\odot}}\right)^{1/3}\left(\dfrac{t}{\rm Myr}\right)^{-1/3}\left(\dfrac{r}{\rm au}\right)^{1-p}M_{\rm E}~{\rm yr}^{-1}.
\label{flux-B}
\end{eqnarray}
When $p=1$, the $r$ dependence is canceled out, and we get Eq. (\ref{flux}), which is exactly the same expression with Eq. (14) of \citet{lam14b}. By this assumption ($p=1$), the dust mass flux is uniform. When we also substitute $t=0.25~{\rm Myr}$ into Eq. (\ref{flux}),
\begin{eqnarray}
\dot{M}_{\rm d}=1.5\times10^{-4}\left(\dfrac{\Sigma_{\rm 1au}}{500~{\rm g~cm}^{-2}}\right)\left(\dfrac{Z_{\Sigma,0}}{0.01}\right)^{5/3}\left(\dfrac{M_{*}}{M_{\odot}}\right)^{1/3}M_{\rm E}~{\rm yr}^{-1},
\label{flux-C}
\end{eqnarray}
which is consistent with the results of our simulations.

By substituting Eq. (\ref{tmig}) for the upper expression of Eq. (\ref{Sigmapl_est}) with Eqs. (\ref{gasg}) and (\ref{temperature}), we get general approximate expressions of the planetesimal surface density:
\begin{eqnarray}
\Sigma_{\rm{pls,est}}
=\dfrac{33.5}{2.728+1.082p}\left(\dfrac{\Sigma_{\rm ump,g}}{500~{\rm g~cm}^{-2}}\right)^{-1}\left(\dfrac{T}{280~{\rm K}}\right)\left(\dfrac{M_{\rm pl}}{20~M_{\rm E}}\right)^{-1} \nonumber \\
\times\left(\dfrac{M_{*}}{M_{\odot}}\right)^{1/2}\left(\dfrac{\dot{M}_{\rm d}}{1.5\times10^{-4}~M_{\rm E}~{\rm yr}^{-1}}\right)\left(\dfrac{r}{\rm au}\right)^{-1}{\rm g~cm}^{-2},
\label{Sigmapl_est-A}
\end{eqnarray}
or
\begin{eqnarray}
\Sigma_{\rm{pls,est}}
=\dfrac{33.5}{2.728+1.082p}\left(\dfrac{\Sigma_{\rm 1au}}{500~{\rm g~cm}^{-2}}\right)^{-1}\left(\dfrac{T_{\rm 1au}}{280~{\rm K}}\right)\left(\dfrac{M_{\rm pl}}{20~M_{\rm E}}\right)^{-1} \nonumber \\
\times\left(\dfrac{M_{*}}{M_{\odot}}\right)^{1/2}\left(\dfrac{\dot{M}_{\rm d}}{1.5\times10^{-4}~M_{\rm E}~{\rm yr}^{-1}}\right)\left(\dfrac{r}{\rm au}\right)^{p-q-1}{\rm g~cm}^{-2}.
\label{Sigmapl_est-B}
\end{eqnarray}
When we substitute $p=1$ and $q=1/2$ into Eq. (\ref{Sigmapl_est-B}), we get the lower expression of Eq. (\ref{Sigmapl_est}).

If we substitute Eq. (\ref{flux-A}) into Eqs. (\ref{Sigmapl_est-A}) and (\ref{Sigmapl_est-B}), we get Eq. (\ref{Sigmapl_est2}) and
\begin{eqnarray}
\Sigma_{\rm{pls,est}}
=\dfrac{33.5}{2.728+1.082p}\left(\dfrac{Z_{\Sigma,0}}{0.01}\right)^{5/3}\left(\dfrac{T_{\rm 1au}}{280~{\rm K}}\right)\left(\dfrac{M_{\rm pl}}{20~M_{\rm E}}\right)^{-1} \nonumber \\
\times\left(\dfrac{M_{*}}{M_{\odot}}\right)^{1/2}\left(\dfrac{t}{\rm Myr}\right)^{-1/3}\left(\dfrac{r}{\rm au}\right)^{-q-1/2}{\rm g~cm}^{-2},
\label{Sigmapl_est-C}
\end{eqnarray}
respectively. Here, $\Sigma_{\rm{pls,est}}$ does not depend on $\Sigma_{\rm g,unp}$ except for the week dependence in the coefficient $33.5/(2.728+1.082p)$, because it is canceled out. When $p=1$ and $q=1/2$,
\begin{eqnarray}
\Sigma_{\rm{pls,est}}
=5.6\left(\dfrac{Z_{\Sigma,0}}{0.01}\right)^{5/3}\left(\dfrac{T_{\rm 1au}}{280~{\rm K}}\right)\left(\dfrac{M_{\rm pl}}{20~M_{\rm E}}\right)^{-1} \nonumber \\
\times\left(\dfrac{M_{*}}{M_{\odot}}\right)^{1/2}\left(\dfrac{t}{\rm Myr}\right)^{-1/3}\left(\dfrac{r}{\rm au}\right)^{-1}{\rm g~cm}^{-2}.
\label{Sigmapl_est-E}
\end{eqnarray}
When we also substitute $t=0.25~{\rm Myr}$, $M_{\rm pl}=20~M_{\rm E}$, and $M_{*}=M_{\odot}$ into Eq. (\ref{Sigmapl_est-E}), we get Eq. (\ref{Sigmapl_est2}).

\section{Gap model used in the cases with the planetary growth} \label{gapmodel}
In Section \ref{realistic}, we use the gap model used in Paper 1 \citep{shi20} to express the cases with the planetary growth. This model is more accurate than that by \citet{duf15a} when the planet is heavy. The perturbed gas surface density is
\begin{equation}
\Sigma_{\rm g}=\Sigma_{\rm g,unp}\max(s_{\rm K}, s_{\rm min}),
\label{Sigma_g_real}
\end{equation}
where $s_{\rm K}=\max(s_{\rm Kepler}, s_{\rm Rayleigh})$. The factor $s_{\rm Kepler}$ is
\begin{equation}
s_{\rm Kepler}=\begin{cases}
\exp\left(-\dfrac{C}{9|x|^{3}K}\right) & (|x|>\Delta) \\
\exp\left(-\dfrac{C}{9\Delta^{3}K}\right) & (|x|\leq\Delta),
\end{cases}
\label{sKepler}
\end{equation}
where $C=0.798$, $\Delta=1.3$, and $x=(r-r_{\rm pl})/H_{\rm g,pl}$. The factor $s_{\rm Rayleigh}$ is
\begin{equation}
s_{\rm Rayleigh}=\begin{cases}
\exp\left(-\dfrac{5}{6}x_{\rm m}^{2}+\dfrac{5}{4}x_{\rm m}|x|-\dfrac{1}{2}x^{2}\right) & (|x|>\Delta) \\
\exp\left(-\dfrac{5}{6}x_{\rm m}^{2}+\dfrac{5}{4}x_{\rm m}\Delta-\dfrac{1}{2}\Delta^{2}\right) & (|x|\leq\Delta),
\end{cases}
\label{sRayleigh}
\end{equation}
where $x_{\rm m}=\{(4/3)CK\}^{1/5}$ is the outer edge of the marginal Rayleigh stable region (i.e., $|x|<x_{\rm m}$). The factor $s_{\rm min}$ is given by \citep{kan15a}:
\begin{equation}
s_{\rm min}=\dfrac{\Sigma_{\rm g,pl}}{\Sigma_{\rm g,unp}}=\dfrac{1}{1+0.04K}.
\label{smin}
\end{equation}

\end{appendix}
\end{document}